\documentclass[12pt,sn-nature]{sn-jnl}
\usepackage[T1]{fontenc}

\usepackage{amsmath}
\usepackage{graphicx}
\usepackage[numbers]{natbib}
\newcommand{\araa}{Annu. Rev. Astron. Astrophys.}   % Annual Review of Astron and Astrophys
\newcommand{\aj}{Astron. J.}   % Astronomical Journal
\newcommand{\apj}{Astrophys. J.}   % Astrophysical Journal
\newcommand{\apjl}{Astrophys. J. Lett.}   % Astrophysical Journal, Letters
\newcommand{\aap}{Astron. Astrophys.}   % Astronomy and Astrophysics
\newcommand{\jgr}{J. Geophys. Res.}   % Journal of Geophysics Research
\newcommand{\mnras}{Mon. Not. R. Astron. Soc.}   % Monthly Notices of the RAS
\newcommand{\nat}{Nature} % Nature
\newcommand{\planss}{Planet. Space Sci.}   % Planetary Space Science
\newcommand{\pasa}{Publ. Astron. Soc. Aust.}   % Publications of the Astron. Soc. of Australia
\newcommand{\solphys}{Sol. Phys.}   % Solar Physics
\title{Coherent radio bursts from known M-dwarf planet host YZ~Ceti}%++++++++++++++++++++++++++++++++++++++++++++++++

\author[1,4,5]{J.\ Sebastian Pineda}
\author[2,3,5]{Jackie Villadsen}
\affil[1]{Laboratory for Atmospheric and Space Physics, University of Colorado Boulder, Boulder, CO, USA}
\affil[2]{Bucknell University, Department of Physics \& Astronomy, Lewisburg, PA, USA}
\affil[3]{Vassar College, Physics and Astronomy Department, Poughkeepsie, NY, USA}

\affil[4]{sebastian.pineda@lasp.colorado.edu}

\affil[5]{these authors contributed equally to this work}

\begin{document}

\flushbottom
\maketitle

\thispagestyle{empty}

%\noindent Please note: Abbreviations should be introduced at the first mention in the main text – no abbreviations lists. Suggested structure of main text (not enforced) is provided below.

%\section*{Introduction}

% Summary Here, $<$200 words ; potential of radio, new detections, tests of emission, stars, conclusions\\
    
\noindent \textbf{Observing magnetic star-planet interactions (SPI) offers promise for determining magnetic fields of exoplanets. Models of sub-Alfv\'enic SPI predict that terrestrial planets in close-in orbits around M~dwarfs can induce detectable stellar radio emission, manifesting as bursts of strongly polarized coherent radiation observable at specific planet orbital positions. We present 2-4~GHz detections of coherent radio bursts on the slowly-rotating M~dwarf YZ~Ceti, which hosts a compact system of terrestrial planets, the innermost orbiting with a 2-day period. Two coherent bursts occur at similar orbital phases of YZ~Cet~b, suggestive of an enhanced probability of bursts near that orbital phase. We model the system's magnetospheric environment in the context of sub-Alfv\'enic SPI and determine that YZ~Ceti~b can plausibly power the observed flux densities of the radio detections. However, we cannot rule out stellar magnetic activity, without a well characterized rate of non-planet-induced coherent radio bursts on slow rotators. YZ~Ceti is therefore a candidate radio SPI system, with unique promise as a target for the long-term monitoring.}

\section*{Main}

%Main text to follow here...(2000-3000 words) 
%\vspace{12pt}

\noindent The possible detection of coherent radio emissions associated with an exoplanetary system has motivated searches from MHz to GHz frequencies because of such emissions' potential to probe the unknown magnetic properties of exoplanets  \citep{Zarka2007,Hallinan2013,Pineda2018}. These proposed emissions are the consequence of a magnetic star-planet interaction (SPI) in which the dissipated energy powers electron cyclotron maser (ECM) emission, which occurs at the cyclotron frequency of the source region: MHz-frequency radiation from the planet itself (fields $< 10$s of G) \citep{Lazio2004,Griessmeier2007}, or MHz-to-GHz radiation from the stellar corona (up to kG fields) as the planetary perturbation is communicated starward via Alfv\'{e}n waves \citep{Saur2013, Turnpenney2018}. The latter mechanism, analogous to the Jupiter-Io flux tube interaction, relies on the host-satellite system being within a sub-Alfv\'enic regime, in which the Alfv\'en speed exceeds the stellar wind speed in the reference frame of the planet.

Based on the example of the Jupiter-Io system \citep{Queinnec1998JGR...10326649Q}, we expect such sub-Alfv\'{e}nic radio SPI to appear as bursts of coherent emission with strong circular polarization lasting minutes to hours. While magnetic interaction can drive near-continuous radiation from the system, the angular beaming of the radio emission from the star-planet flux tube, as viewed by a distant observer, should cause the emission to appear as well-defined bursts dependent upon the satellite orbital phase.  

Recent results revealed 150-MHz ECM emission from M-dwarf stars that could be consistent with sub-Alfv\'{e}nic SPI \citep{Vedantham2020, Callingham2021NatAs...5.1233C, Pope2021ApJ...919L..10P}. However, these systems need confirmation that a planetary satellite indeed drives the radio emission, with no planets yet found in campaigns targeting GJ~1151 \citep{Pope2020,Perger2021}. Polarized radio emission from Proxima Centauri exhibits possible orbital periodicity with Prox Cen b \citep{PerezTorres2021AA...645A..77P}, but the planet's 11-day period places it at an orbital distance unlikely to have sub-Alfv\'enic interaction \citep{Kavanagh2021MNRAS.504.1511K}. 
Moreover, the possibility of coherent radio bursts entirely of stellar origin remains a significant possibility as magnetically active M-dwarf stars frequently exhibit polarized radio emissions \citep{Lynch2017,Villadsen2019}, and the slowly rotating M-dwarf Prox Cen exhibits stellar flare-associated coherent radio bursts \citep{Zic2020ApJ...905...23Z}. The radio flaring properties of inactive M-dwarfs across MHz to GHz frequencies are largely unknown, complicating efforts to exclude stellar activity as a cause of radio bursts. To disentangle stellar activity and SPI, we aim to identify a system with coherent radio bursts and a very short period planet (less than a few days), which will enable long-term monitoring to test orbital periodicity --- the clear evidence that could conclusively determine that any emissions are powered by SPI.

In this work, we report the detection of 2-4~GHz coherent radio bursts from the known exoplanet host YZ~Ceti using \textit{NSF's Karl G. Jansky Very Large Array} (VLA) \citep{Perley2011}. This nearby slowly-rotating star has three small planets orbiting in a compact configuration \cite{AstudilloDefru2017,Stock2020}, including one in a 2-day period. We discuss our radio observations in the context of star-planet interactions, considering whether the YZ~Ceti planets could plausibly power the detected polarized bursts, and whether their recurrence suggests a solely stellar or possibly planet-induced origin.

\begin{figure*}[tbp]
	\centering
	\includegraphics{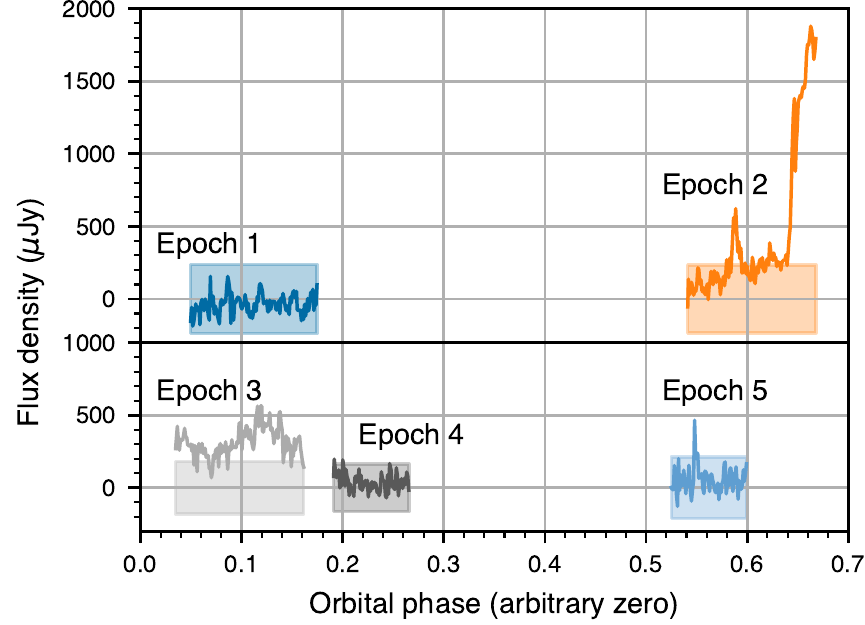}
	\caption{Time series of all 2-4~GHz observations of YZ~Cet, phased using YZ~Cet~b's orbital period, binned to 3-minute intervals. The first two epochs are displayed on a separate panel to avoid overlap. The shaded regions show $\pm$3 times the estimated error on the flux density for a 3-minute time bin in each epoch, so that points above the shaded region have $>3\sigma$ significance. The coherent burst in Epoch 2 (phase $\sim$0.59) does not recur at the same orbital phase in Epoch 5. The time error on phase-wrapping between Epochs 1 to 5 is 5.3 minutes, negligible on the scale of this plot.\label{fig:phased_tseries}}
\end{figure*}

\noindent \textbf{Characterizing the radio bursts.} We observed YZ~Ceti at 2-4~GHz with the VLA in 5 epochs: an initial program of three daily 6.5-hour observations on 30 Nov to 2 Dec 2019, and two 4-hour follow-up observations on 2 Feb and 29 Feb 2020. Accounting for calibrator observations our total on source time is $\sim$26 hours. Figure~\ref{fig:phased_tseries} shows the time series of all five epochs, phase-wrapped to the 2.02087~d orbital period of inner planet YZ~Cet~b \citep{Stock2020}, with an arbitrary zero phase because the radial velocity-determined orbital phase has an error of $\sim$1/8 of the orbital period, too large to check whether bursts occur at quadrature.  In the initial observations, the system was undetected ($<36$~$\mu$Jy) in Epoch 1, emitted multiple radio bursts in Epoch 2, and produced slowly variable quiescent emission ($313\pm20$~$\mu$Jy) in Epoch 3.  In the follow-up observations, the star was undetected ($<100$~$\mu$Jy) in Epoch 4, and a single coherent burst occurred in Epoch~5.

\begin{figure}[tbp]
	\centering
    \includegraphics[width=\textwidth]{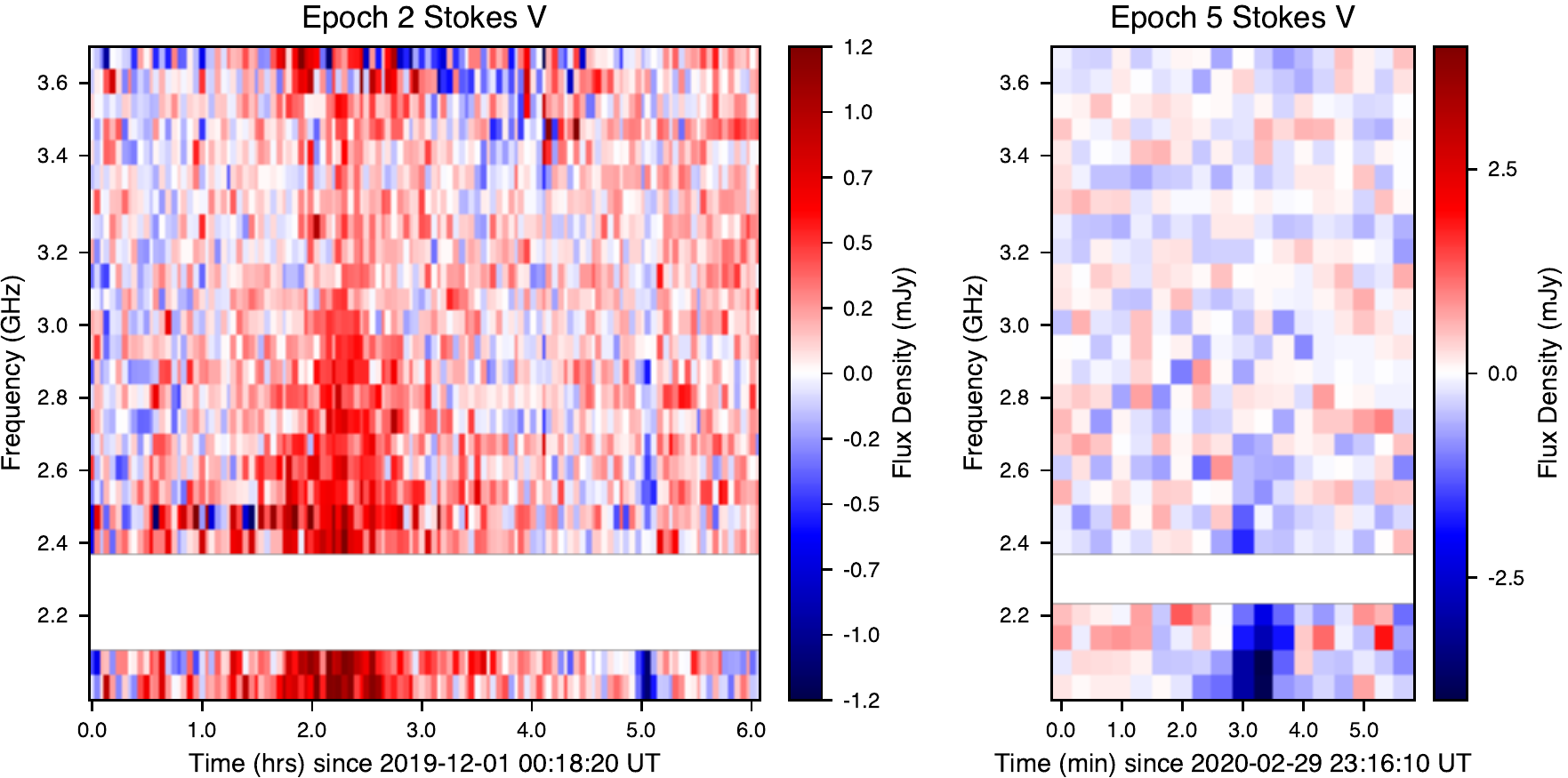}
	\caption{2-3.75~GHz Stokes V dynamic spectra of detected bursts. Right-polarized bursts appear in red and left-polarized in blue. \textit{(Left)} All 6.5 hours of Epoch~2, with an uncertainty of 195~$\mu$Jy in each pixel of 3 minutes and 64 MHz. Above $\sim$3.4~GHz the noise increases due to RFI. \textit{(Right)} 6-minute excerpt of Epoch~5, with an uncertainty of 520~$\mu$Jy in each pixel of 20 sec and 64 MHz.  The events at 2.3 and 5.1 hours in Epoch 2, and the Epoch 5 event, share a declining spectrum with the brightest emission observed near 2~GHz.  All three bursts have peak flux densities of $>5\sigma$ in the dynamic spectrum.\label{fig:dyn}}
\end{figure}

Figure~\ref{fig:dyn} (left panel) shows the Stokes V dynamic spectrum of Epoch~2 (see Methods for polarization time-series). An hour-long coherent burst, with nearly 100\% right circular polarization (RCP), occurs with peak flux density of 620$\pm$80~$\mu$Jy (in the Stokes~I time series) around 2.3~hours after the start of the observation.  Three hours later, a bright flare occurs with weak polarization, which favors the incoherent gyrosynchrotron mechanism responsible for many solar and stellar flares \citep{Gudel2002ARAA..40..217G}. The flare is preceded by enhanced Stokes I emission from 3 to 5 hours, with weak right polarization, which may be due to slowly-varying quiescent emission (consistent with the variable quiescent emission in Epoch 3) or pre-flare activity. An additional small burst, whose strong left circular polarization (LCP) favors a coherent mechanism, occurs during this epoch at 5.1 hours, where it is superimposed on the incoherent flare.
This LCP event may be due to accelerated electrons during the impulsive phase of a flare \citep[e.g.,][]{Zic2020ApJ...905...23Z}. Since this LCP feature is likely associated with stochastic stellar flaring, we do not consider it further here. In our discussion, we focus on the RCP burst at 2.3 hours, since we are searching for coherent bursts induced by SPI.

We infer a coherent emission mechanism based on degree of polarization; for non-thermal processes, incoherent gyrosynchrotron emission with harmonic $s\sim10-100$ has less than 60\% polarization for most viewing angles \citep{Dulk1985ARA&A..23..169D} whereas coherent emission often has up to 100\% polarization.  At GHz frequencies, without a measured source size, the brightness temperature indicates a non-thermal mechanism for both the strongly polarized bursts and weakly polarized flare and quiescent emission.  For example, the 620~$\mu$Jy peak flux density of the RCP burst at 3~GHz corresponds to a brightness temperature of $>1.5 \times 10^9$~K for an upper limit on source size of the full stellar disk; GHz-frequency coherent sources are likely much smaller, tracing to individual magnetic footpoints in the stellar corona. With the evidence available, we cannot differentiate between two possible coherent emission mechanisms, plasma emission and electron cyclotron maser, the latter of which is expected for SPI.
An ECM mechanism for YZ~Cet's coherent bursts is plausible because many other M dwarf radio bursts have been attributed to ECM due to high brightness temperature \citep{Osten2008ApJ...674.1078O,Vedantham2020,Callingham2021NatAs...5.1233C} or x-mode polarization \citep{Villadsen2019}, however we do not rely on the emission mechanism to assess an SPI origin.
Instead, we search for evidence of orbital modulation of bursts to test the possibility that star-planet interaction drives the observed coherent bursts.

We scheduled the Epoch~5 follow-up observations to encompass the same orbital phase as the RCP coherent burst in Epoch~2.  We detected a one-minute-long left-polarized coherent burst (Figure~\ref{fig:dyn}, right panel) in these follow-up observations, with a peak flux density of 465$\pm$70~$\mu$Jy in the Stokes~I time series with 3-minute time bins (using 3 minutes for a consistent comparison to Epoch~2). The orbital phase of the follow-up burst detection does not exactly match that of the observed Epoch~2 bursts (Figure~\ref{fig:phased_tseries}), but instead occurs about 2~hrs earlier in the 2-day orbital period. Below, we consider whether the flux density and the relative timing of the bursts may be consistent with an SPI mechanism.

\noindent \textbf{Assessing star-planet interactions.} The detection of multiple coherent radio bursts from YZ~Ceti prompts the question of whether the planets in the system could have powered the radio bursts. To answer this question we must estimate the magnetized environment of YZ~Ceti, and calculate the strength of the potential star-planet interaction.

For the environment, we adopted an isothermal stellar wind model for the YZ~Ceti system. The planets possess orbital distances ($a/R_{s} \gtrsim 20$)\citep{Stock2020} unlikely to be encompassed by closed stellar magnetic field lines, and thus the planets are likely intersecting open field lines that carry the stellar wind.  
We employ two fiducial wind models (see Methods). Model A assumes an open magnetic field entrained by a strong radial wind launched from the stellar surface, matching assumptions commonly used in approximate calculations in the literature \citep[e.g.,][]{Turnpenney2018,Vedantham2020}. Model B uses a weaker wind and a PFSS extrapolation to account for closed field near the star. Incorporating a closed field region near the stellar surface, in particular, should provide a more realistic estimate of the radial field decay of any magnetized star. Uncertain model assumptions, regarding stellar mass loss rate and magnetic field strength, can impact whether a planet orbits in the sub- or super-Alfv\'enic regime. Encouragingly, both of our fiducial models find that the innermost planets are within the sub-Alfv\'{e}nic regime, allowing the planetary perturbation of the stellar magnetic field to communicate energy back towards the stellar surface to induce GHz emission. We expand on the effects of exploring the wind-model parameter space in the Methods.

To compute the power available to drive planet-induced radio emissions, we employed the frameworks of \cite{Lanza2009} and \cite{Saur2013}. The former (Reconnection) computes the energy released through magnetic reconnection from the obstacle-field interaction, while the latter (Alfv\'{e}n Wing) focuses specifically on the energy in Alfv\'{e}n waves propagating back to the host star from the same interaction. This computation depends on all of the magnetic environment variables defined in our wind models A and B, as well as on the magnetic properties of the planets, with stronger planetary magnetic fields carving out a planetary magnetosphere that serves as the enlarged obstacle radius for the SPI (see Methods). Within both frameworks, and using both models A and B for the magnetized environments, we computed the possible strength of ECM radio bursts associated with YZ~Ceti~b, the closest-in planet and most likely to drive appreciable radio emissions. We further varied the assumed planetary dipole magnetic field strength, and the assumed planet radius to assess their impact on the potential radio burst flux densities.

\begin{figure*}[tbp]
	\centering
	\includegraphics[width=\textwidth]{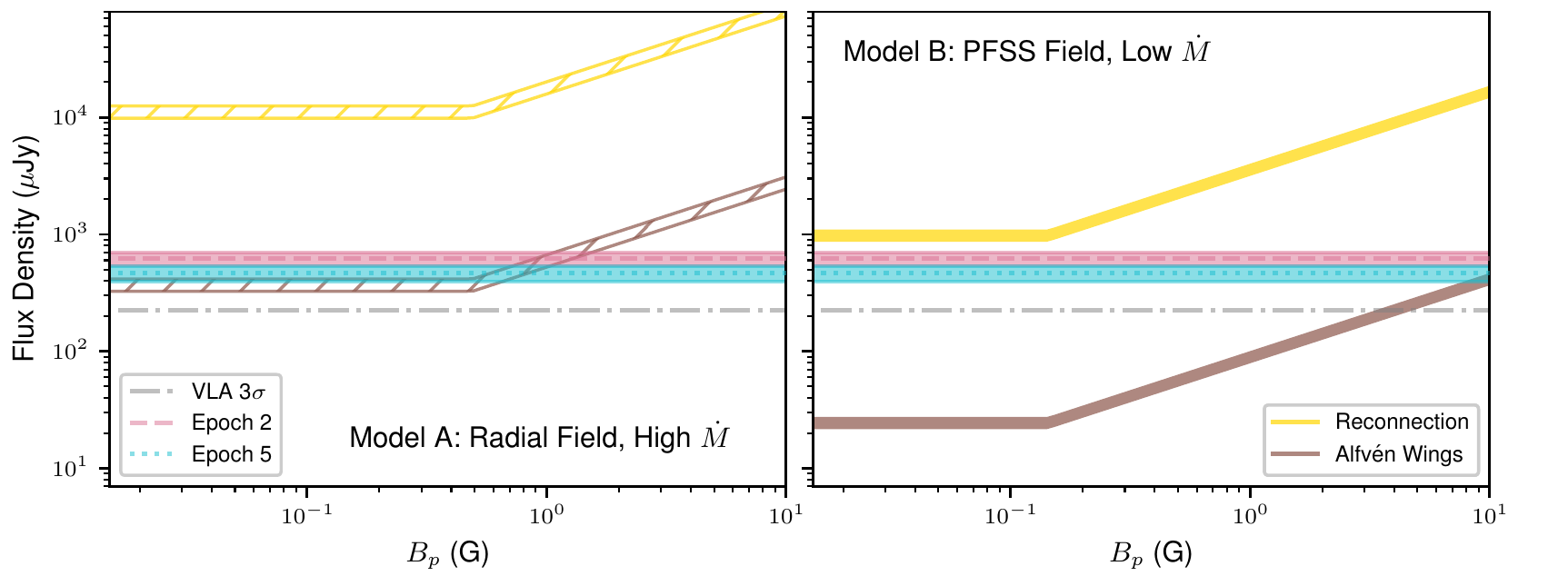} % requires the graphicx package
	\caption{Predicted radio flux density due to star-planet interaction with YZ~Ceti b.  Using our fiducial wind models for the magnetized environment of the YZ~Ceti system, A (\textit{Left}) and B (\textit{Right}), we predict radio flux densities generated by the planet YZ~Ceti~b, interacting with its host, according to the two interaction frameworks, Reconnection \cite{Lanza2009} (gold) and Alfv\'{e}n Wing \cite{Saur2013} (brown). For a given planet dipole magnetic field, the vertical 
    extent of either SPI framework swath (Reconnection or Alfv\'{e}n Wing) corresponds to the flux density across the range of plausible planet radii (see Methods).
	The Reconnection \citep{Lanza2009} predictions are stronger than those of Alfv\'en Wing \cite{Saur2013}. Our measured flux densities of the brightest coherent radio bursts from Epoch 2 and Epoch 5 are also shown (pink and blue) with shaded regions corresponding to 1$\sigma$ uncertainties. The horizontal grey dash-dot line shows the typical 3$\sigma$ RMS sensitivity in our VLA 3~min time series.}
 \label{fig:fluxden}
\end{figure*}

We demonstrate our results in Figure~\ref{fig:fluxden}, with Model A on the left and Model B on the right, and different swathes for the Reconnection and Alfv\'{e}n Wing predictions in each panel. Under Model A, the Alfv\'{e}n Wing predictions neatly match the measured flux density of the bursts for a weakly magnetized planet ($\sim$1~G); whereas, the Reconnection framework strongly overpredicts the measured bursts. This dramatic difference is likely due to Model A overestimating stellar field strengths at the location of the planet.
Under the Model B paradigm, in which PFSS provides more realistic near-surface radial field decay, YZ~Ceti~b could power the emission with the Reconnection framework, in the absence of an intrinsic planetary magnetic field. However, according to the Alfv\'{e}n Wing scenario, the planet would likely require a strong field ($\gtrsim$ a few~G) to power the detected bursts.

There are many uncertain assumptions that go into these SPI flux predictions, which in effect allow the prediction curves of Figure~\ref{fig:fluxden} to move up and down by factors of several. Nevertheless, following our best characterization of the system (Model B), our calculations suggest that if the Reconnection framework is an accurate description of the physics, then the bursts could readily be produced by YZ~Ceti~b through star-planet interaction. Interestingly, if the Alfv\'{e}n Wing scenario is more applicable to these systems then the radio detections would imply a substantial planetary magnetic field for the terrestrial planet.

After modeling the flux density of the detected radio bursts, we also considered their relative timing in order to evaluate their potential SPI nature. In Jupiter, the observed recurrence of the Io-induced radio emissions depends on both the orbital period and the rotation period of the tilted Jovian magnetic field (see within \citep{Zarka2018A&A...618A..84Z}). These define the synodic period ($P_{\mathrm{syn}} = [P_{\mathrm{orb}}^{-1} - P_{\mathrm{rot}}^{-1} ]^{-1}$) at which the satellite orbit returns to the same position relative to the host magnetic field. \cite{Fischer2019ApJ...872..113F} discussed the possible SPI periodicities in depth, noting the importance of the synodic period and half-synodic period, the latter defining a similar satellite position, but on the opposite side of the host magnetic field.  

\begin{figure*}[tbp]
	\centering
	\includegraphics[width=\textwidth]{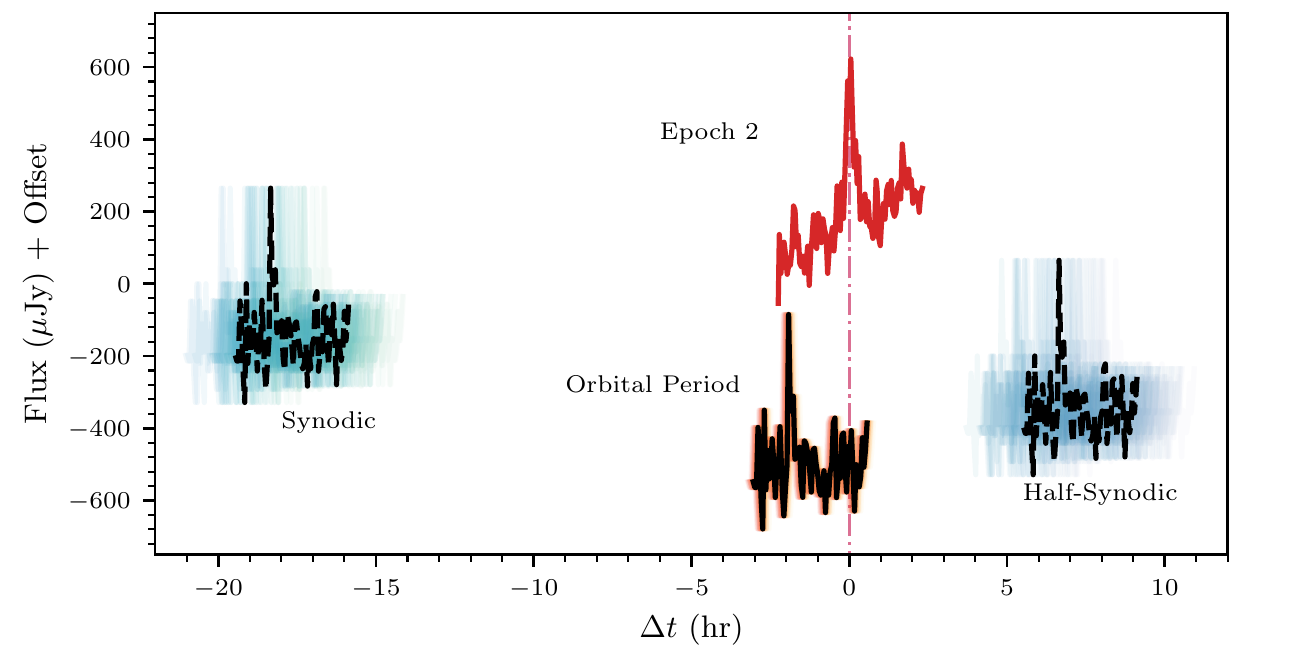} % requires the graphicx package
	\caption{Phase-wrapped time series of the portion of Epoch 2 containing the RCP coherent burst, and all of Epoch 5.  By phasing the Epoch 5 flux density light curves (black lines) at different periods (orbital period, synodic period, and half-synodic period), as compared to the Epoch 2 burst (red, with vertical offset for clarity), we tested whether the radio burst occurrence is related to the orbit of YZ~Ceti~b. The Epoch 5 burst does not show a perfect recurrence at the orbital period ($\Delta t$=0), but instead a difference in timing of $\sim$2 hrs, and $\sim$6 hrs when using the half-synodic period. The burst phasing using the synodic period indicates a burst separation of almost a half-period; a full period is just over 48 hrs. The semi-transparent colored lines (blue-green/purple/orange) below the Epoch 5 curves illustrate the phase error due to uncertainty in the period measurements.}
	\label{fig:phasing}
\end{figure*}

We phase-wrapped the time series of the RCP burst from Epoch~2 and the time series of Epoch~5, using three different relevant periods for YZ~Ceti~b (see Figure~\ref{fig:phasing} in Methods): orbital (2.02087~d), synodic (2.08232~d), and half-synodic (1.04116~d). To calculate the synodic period, we used the photometric rotation period from \cite{Stock2020}, $68.46 \pm 1.00$, leading to uncertainties of order an hour in the synodic and half-synodic phase-wrapping, whereas orbital period wrapping is precise to within a few minutes. When phase-wrapping with the orbital period, the Epoch 5 burst takes place $\sim$2~hr prior to the phase of the Epoch 2 polarized burst (phase difference, $\Delta \phi \sim0.04$). Neither the synodic or half-synodic period phases the Epoch 5 bursts closer to the time of the Epoch 2 burst (see Methods). These misalignments mean that our data cannot provide definitive evidence of SPI. We also applied a similar analysis to planets c and d, finding worse burst agreement in both time and phase. That there is near orbital recurrence for planet b, however, is tantalizing, since precise burst timing can depend on the complexities of magnetic geometry. 

Changes in the near surface magnetic field could impact the radio beaming angles and influence the observed timing of any planet-induced radio emissions --- the importance of non-dipolar stellar field components increases closer to the star. The fact that the burst recurrence is closer in time to the half-synodic phasing than the synodic phasing also suggests that the Epoch 5 burst occurs on the opposite side of the magnetic field than the Epoch 2 burst, with the emission emerging from different poles. This point is consistent with the change in polarity, RCP vs.\ LCP, between the Epoch 2 and 5 bursts (see Figure~\ref{fig:dyn}).  The difference in source origins with respect to the stellar magnetic field may also be responsible for the change in burst duration between epochs. If we assume the bursts are a consequence of the radio cone sweeping across the line of sight at a rate tied to the planet orbit, then the 1~hr Epoch 2, and 1~min Epoch~5 durations would correspond to cone thicknesses of $7.4^{\circ}$, and $0.12^{\circ}$, respectively. However, even when assuming intrinsically narrow (1$^{\circ}$) cone widths, models of SPI radio dynamic spectra exhibit a wide range of burst durations depending on the extended source geometry, polarity, and viewing angles \citep{Hess2011AA...531A..29H}. A detailed map of the magnetic field structure (e.g., from ZDI) would enable confirmation of these behaviors and further apply consistency checks on the SPI scenario.

The polarized bursts may also be a consequence of ordinary stellar magnetic activity, such as flares, despite the star's slow rotation ($\sim68$~d). If we consider each of the polarized bursts to be independent stochastic events, we can consider our rate of detection as 2 events in 26 hrs (0.0769 hr$^{-1}$) of radio monitoring (neglecting the small LCP events in Epoch 2 since it may be flare substructure). Using simple Poisson statistics, the probability of seeing at least one event in the 3.6~hrs associated with Epoch~5 is $\sim24$\%. On the other hand, the probability of seeing one burst, on two separate occasions within 2 hours of a given phase (a 4~hr window) is only 5.1\% --- small but insufficient as conclusive evidence. It is thus plausible that the bursts could have no association with the planetary system and be a normal part of the radio stellar activity of slowly rotating M dwarfs, which is not yet well-studied. The mechanisms powering the emission remain inconclusive, and we thus categorize YZ~Ceti as a SPI candidate, requiring further follow-up to discern the nature of the radio bursts.

\noindent \textbf{Conclusions.} We detected 3-GHz coherent radio bursts from the YZ~Cet system, occurring in two of five observed epochs, where the coherent emission mechanism was indicated by a non-thermal brightness temperature and high degree of circular polarization. The 3-GHz frequency is consistent with ECM emission originating from the kG fields expected at magnetic footpoints in the low stellar corona. The coherent radio bursts in two epochs nearly recurred in phase with the orbital period of YZ~Cet~b, after 90.9 days. The flux density of these bursts is roughly consistent with predictions for the average luminosity of sub-Alfv\'enic SPI \citep{Lanza2009,Saur2013}, depending on the assumed conditions of the stellar environment and planetary magnetic field. Based on their luminosity and proximity in orbital phase, we consider these two events as candidate SPI events, but cannot rule out stellar magnetic activity as a possible cause. If SPI in this system is confirmed, the radio luminosity will enable estimation of the magnetic field strength of YZ~Cet b, particularly when combined with refined measurements of the stellar magnetic field and the theory development of accurate flux density predictions.

%For the possibility of star-planet interaction, we considered both the luminosity and orbital phase of the bursts.  The flux density of these bursts is roughly consistent with predictions for the average luminosity of sub-Alfv\'enic SPI \citep{Lanza2009,Saur2013}, depending on the assumed conditions of the stellar environment and planetary magnetic field. If SPI in this system is confirmed, refined measurements of the stellar field environment and development of the dominant applicable theoretical framework would enable a clear magnetic characterization of the planet YZ~Ceti~b. 

The bursts do not recur at an exactly consistent orbital phase in follow-up observations (phase difference, $\Delta \phi \sim0.04$), and display a greater degree of phase separation when considering the synodic and half-synodic periods, which take into account the rotation of the star. Recurrence with an orbitally dependent phase would provide a direct confirmation of planet-induced radio emission. In its absence, the possibility of orbital modulation remains: the preferred orbital phase for Io-induced bursts from Jupiter changes between two values over the course of Jupiter's rotational period \citep[e.g.,][]{Zarka2018A&A...618A..84Z}, due to the tilt of Jupiter's magnetic field. If the star's dipole field is tilted, then it rotated significantly between Epochs 2 and 5, since their $\sim$90-day separation is 1.3 times the star's $\sim$68-day rotational period. In spite of this ambiguity, the result is suggestive: the orbital phase difference is small, and the periodicity may reflect an orbit-dependent window of visibility \citep{Saur2013} for observing the beamed radio emission. Further observations that can test for orbital modulation include longer-term radio monitoring and spectropolarimetric observations to determine the orientation of the star's large-scale magnetic field.

Monitoring to search for orbital modulation must contend with the ``foreground'' of events caused by ordinary stellar activity. YZ~Ceti's slow rotation period places it among stars with weak magnetic activity \citep[e.g.,][]{Newton2017ApJ...834...85N}. Slow rotators cannot power their luminous radio bursts by corotation breakdown in circumstellar plasma \citep{Nichols2012ApJ...760...59N,Vedantham2020}, a process responsible for some non-Io radio bursts from Jupiter. However, slowly-rotating M~dwarfs can still release energy through magnetic reconnection to produce luminous flares at other wavelengths such as UV \citep[e.g.,][]{Loyd2018ApJ...867...71L}. Slow rotator Prox~Cen has also produced coherent radio bursts near GHz frequencies \citep{Slee2003}, including an optical flare-associated event \citep{Zic2020ApJ...905...23Z},
%\citep{Slee2003,Zic2020ApJ...905...23Z}, with unclear planetary association \citep{PerezTorres2021A&A...645A..77P,Kavanagh2021MNRAS.504.1511K},
suggesting that slowly rotating M~dwarfs are capable of coherent bursts due to stellar activity. On the Sun, coherent bursts driven by magnetic reconnection are sometimes associated with incoherent gyrosynchrotron flares \citep{Dulk1985ARA&A..23..169D}; similarly, the coherent bursts and incoherent flare in Epoch 2, occurring over the span of 4 hours, may all derive from related processes in a magnetically active region. 

A deeper understanding of polarized stellar radio bursts (rates, morphology, physical drivers) across MHz-GHz frequencies would provide a significant advancement in disentangling such emissions from potential SPI signals. Searches for SPI emissions will need to contend with this foreground of stellar activity to be successful. In light of these considerations, and our own candidate detections, we propose a general criteria for assessing and confirming magnetic SPI at radio frequencies. The conditions are two-fold: 1) recurrence of radio bursts at a period dependent upon the orbit of a confirmed planet, and 2) a Poisson probability $p<0.0027$ (equivalent to 3$\sigma)$ of randomly observing those events within a narrow phase/time window, where that probability is based on an average burst rate determined by observing a wide range of orbital phases. If we again assume a stochastic rate of, 0.0769~hr$^{-1}$, we would require 4 phased bursts within a 4-hour phase window to exceed this probability threshold; at 5$\sigma$ confidence, we would require 10 such bursts. It is prior knowledge of the planet's period to high precision that can enable high confidence in the SPI interpretation given a low probability of randomly recurring stochastic events. Radio non-detections, spanning a broad phase range, are important for an accurate measurement of a stochastic radio burst rate in order to distinguish the phases of SPI enhancement from standard stellar processes.
These criteria rely largely on multi-epoch radio monitoring, and can be corroborated by complementary observations of the stellar magnetic field and planetary geometry. For such long-term monitoring to test periodicity, YZ~Ceti's confirmed planet in a 2-day period makes it a uniquely promising case study for magnetic star-planet interactions.

\section*{Methods}

\begin{table}[ht]
\begin{center}
\caption{Slowly Rotating M-dwarfs with Coherent Radio Detections
\label{tab:prop}}
\begin{tabular}{l c c c}
\hline
Property  & YZ Ceti & Proxima Centauri & GJ~1151 \\
\hline
Spectral Type & M4.5 \citep{Reid1995} & M5.5 \citep{Hawley1996} & M4.5 \citep{Reid1995} \\
Distance (pc) & $3.712$ &   $1.301$  & $8.038$ \\
Mass ($M_{\odot}$) & $0.137 \pm 0.003$ & $0.123 \pm 0.003$ & $0.168 \pm 0.004 $ \\
Radius ($R_{\odot}$) & $ 0.163 \pm 0.007$ & $0.147 \pm 0.005$ & $ 0.192 \pm 0.008 $ \\
$L_{\mathrm{bol}}$ ($10^{30}$ erg s$^{-1}$) & $ 8.6 \pm 0.1 $ & $6.00 \pm 0.08$ & $ 13.0 \pm 0.2 $ \\
$T_{\mathrm{eff}}$ (K) & $3110 \pm 70 $& $2990 \pm 50$& $3181 \pm^{65}_{63} $ \\
$\log_{10} \, [L_{\mathrm{X}} / L_{\mathrm{bol}}]$ & $-4.13$ \citep{Stelzer2013} & $-4.4$  \citep{Wargelin2017MNRAS.464.3281W} & $<-4.19$ \citep{Stelzer2013} \\
$\log_{10} \, [L_{\mathrm{H}\alpha}/ L_{\mathrm{bol}}]$ & $-4.32$ \citep{Reiners2018} & $-4.20$ \citep{Hawley1996} & $-4.75$ \citep{Reiners2018}\\
$P_{\mathrm{rot}}$ (d) & $\sim$68.5 \citep{Stock2020} & $\sim$89 \citep{Newton2018} & $\sim$130 \citep{Irwin2011} \\
$Bf$ (kG) & $\sim$2.2 \citep{Moutou2017} & $\sim$0.6 \citep{Reiners2008b} & ---  \\
Planetary System & Y & Y & ? \citep{Pope2020,Perger2021} \\
Planet Periods (d)  & 2.02, 3.06, 4.66 \citep{Stock2020} & 5.1243$^{\dag}$ \citep{Faria2022AA...658A.115F}, 11.186 \citep{AngladaEscude2016} & --- \\
Planet $a/R_{s}$& 21.6, 28.4, 37.6 \citep{Stock2020} & 42.2$^{\dag}$ \citep{Faria2022AA...658A.115F}, 70.95 \citep{AngladaEscude2016}  & --- \\
Planet $m_{p} \sin i$ ($M_{\oplus}$)  & 0.70, 1.14, 1.09 \citep{Stock2020} & 0.40$^{\dag}$ \citep{Faria2022AA...658A.115F}, 1.27 \citep{AngladaEscude2016} & --- \\
\hline 
Frequency (GHz) & 3.0 & 0.9-2.0 & 0.145 \\
$L_{\nu,iso} {^*}$ (erg s$^{-1}$ Hz$^{-1}$) & $10^{13}$ & 6-100 $\times 10^{12}$ \citep{Slee2003,Zic2020ApJ...905...23Z,PerezTorres2021AA...645A..77P} & $10^{14}$ \citep{Vedantham2020} \\
%The 10 for YZ~Cet Liso is calculated from 620 $\mu$Jy peak flux for RCP burst
Burst Duration (hrs) & 1 &  $0.07 - 38.4+$  & $>$ 8  \\
Polarization (\%) & $\sim$93 & $\sim90-100$ &  $\sim64$   \\
\hline

\multicolumn{4}{p{\linewidth}}{ \small $\dag$ The quoted innermost planet for Prox.\ Cen.\ is a new candidate from \cite{Faria2022AA...658A.115F}. }\\

\multicolumn{4}{p{\linewidth}}{ \small * $L_{\nu,iso} = 4\pi d^2 S_\nu$ assumes isotropic radiation, where $S_\nu$ is the approximate peak flux density of an event in this work (YZ~Ceti) or cited works. This assumption is not accurate for coherent emission, but it provides a distance-independent measure for comparing stars without relying on uncertain estimates of the true beaming angle.} 

\end{tabular}
\end{center}

\end{table}

\noindent \textbf{Stellar Characterization.} Table~\ref{tab:prop} details physical stellar properties for our target YZ~Ceti and comparison objects Prox.\ Cen.\ and GJ~1151, determined by combining multiple empirical relations to jointly constrain the mass, radius, and, bolometric luminosity \citep[method in][]{Pineda2021ApJ...918...40P}. The estimates employ a precise parallax from \textit{Gaia DR2} \citep{Gaia2018A&A...616A...1G}, with the effective temperature derived from combining the luminosity and radius. YZ~Ceti and GJ~1151 have bolometric flux measurements from \cite{Mann2015}, and for Prox.\ Cen.\ it is taken from \cite{Boyajian2012}. We further quote additional relevant activity properties, including the stellar rotation period ($68.46\pm 1.00$), from their respective references. There is a more precise V-band rotation period of $68.4 \pm 0.05$ \citep{Stock2020} measured, but the formal statistical error may not encompass all forms of systematic error. To explore the impact of rotation period uncertainty over the $\sim$90 day separation between Epoch 2 and 5, we opted to use the less precise period when phase wrapping with the synodic periods, finding that it broadens the uncertainty curves but does not impact our conclusions. We also note that optical flaring can be seen in each object's \textit{Transiting Exoplanet Survey Satellite} \citep{Ricker2015} light curves.

\begin{figure}[tbp]
	\centering
	\includegraphics[width=0.8\textwidth]{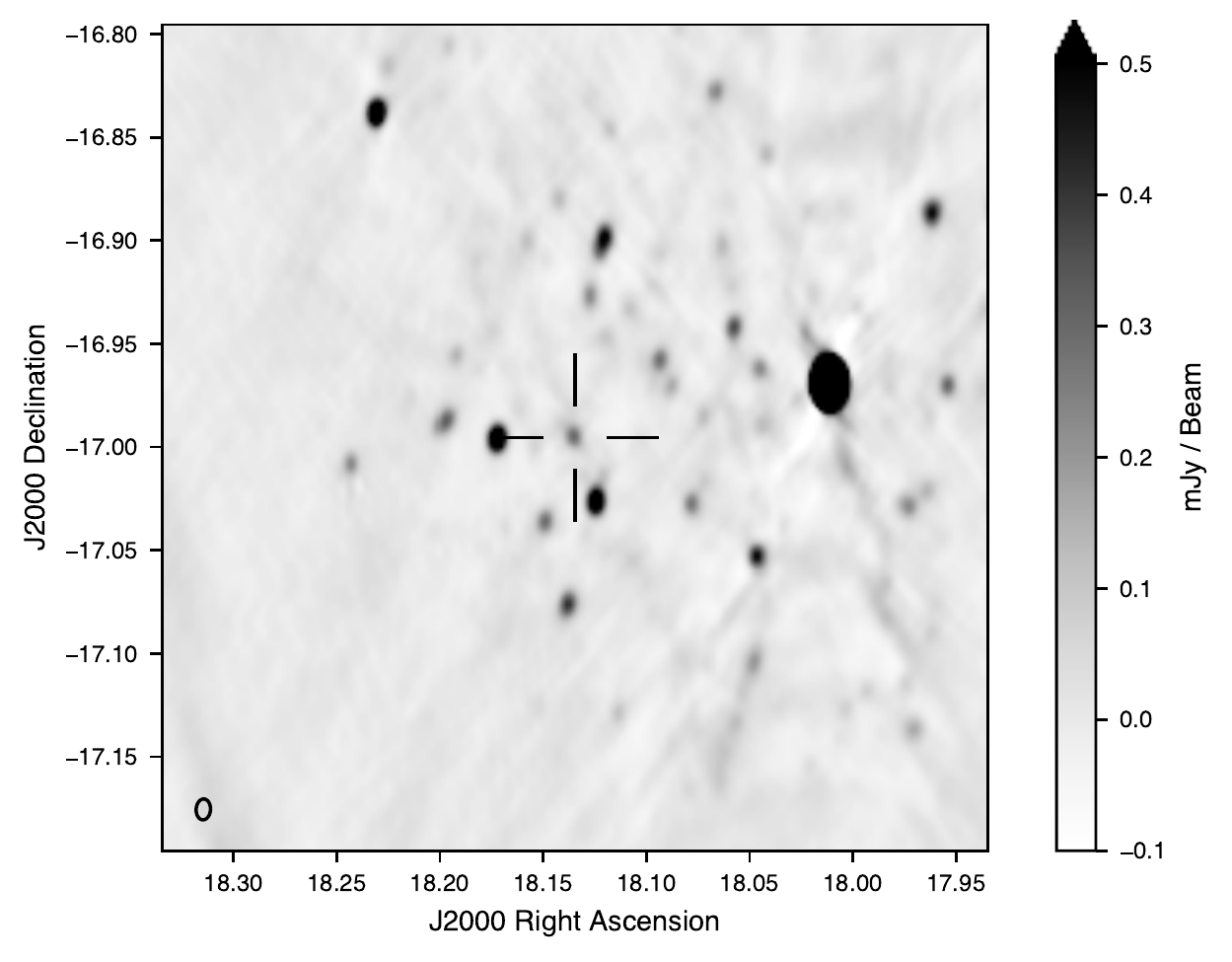}
	\caption{2-4~GHz image of the field surrounding YZ~Cet on 30 Nov to 2 Dec 2019. The crosshairs show the location of YZ~Cet.  The brightest background source, PMN J0112-1658, has a flux density of 150 mJy. YZ~Cet's flux density in this image, an average of Epochs 1 to 3, is 301$\pm$25~$\mu$Jy. \label{fig:fov}}
\end{figure}

\noindent \textbf{Radio Data Analysis.} For our VLA observations of YZ Cet we used 3C147 as flux calibrator and J0116-2052 as gain calibrator, and calibrated the data in CASA \citep{mcmullin2007} using the VLA pipeline. The VLA was in compact configuration: D in Epochs 1-3, DnC in 4, C in 5. We observed with the phase center located halfway between YZ~Cet and the 150-mJy nearby source PMN~J0112-1658 (7.5~arcmin away from YZ~Cet), to keep that source within the main lobe of the primary beam, then shifted the phase center to the location of the star before imaging. We imaged the first three epochs together and each of the two follow-up epochs separately, using CASA's \textbf{tclean} task with W-projection, multiscale imaging, and multifrequency synthesis with 3 Taylor terms.  We used natural weighting to maximize point source sensitivity.

For each data set, we performed self-calibration on the target field using CASA's \textbf{gaincal} command: 1-2 rounds of phase-only self-calibration, followed by 1-2 rounds of amplitude and phase self-calibration. For amplitude self-calibration, we used \textbf{gaincal}'s \texttt{solnorm=True} parameter, normalizing gain amplitudes to an average value of one to avoid artificially increasing flux density. For example, in the self-calibrated image of Epochs 1 to 3 (Figure~\ref{fig:fov}), the peak flux density of PMN~J0112-1658 remained consistent (148.4~mJy before self-calibration, 148.5~mJy after), while the uncertainty significantly improved: the RMS in a region near YZ~Cet that is empty of bright sources was 120~$\mu$Jy before, and 25~$\mu$Jy after self-calibration.

\begin{figure}[tbp]
	\centering
    \includegraphics[width=\textwidth]{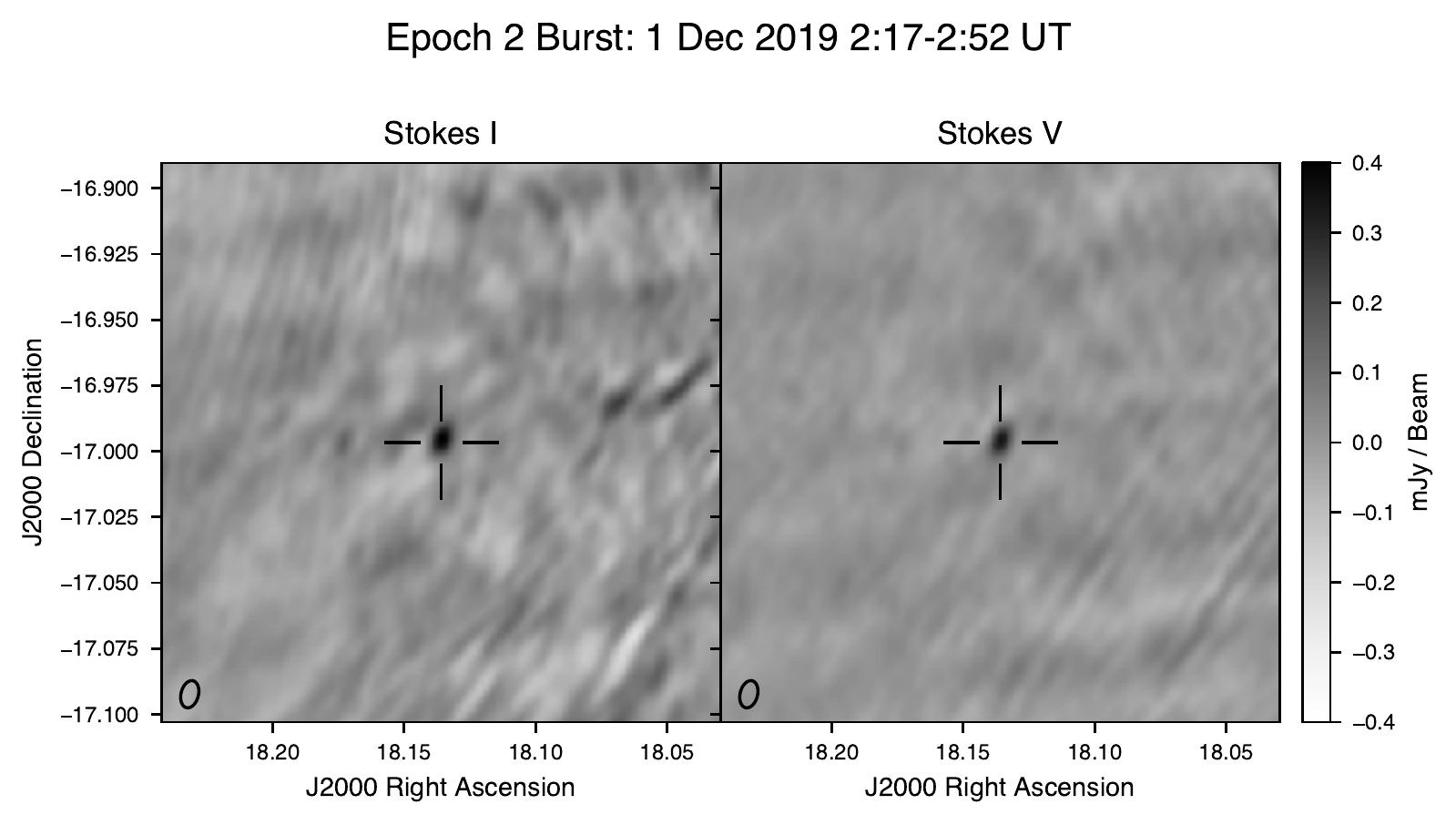}
	\caption{2-4~GHz image of YZ~Cet in Stokes~I \textit{(left)} and Stokes~V \textit{(right)} during the right-polarized coherent burst in Epoch~2. We subtracted background sources from the visibilities before making this image, although some imperfectly removed sidelobes remain in Stokes~I.  We imaged times corresponding to the full-width half max of the burst in the time series, measuring an average flux density of 399$\pm$47~$\mu$Jy (I) and 341$\pm$25~$\mu$Jy (V) for YZ~Cet during this interval. \label{fig:epoch2burst_image}}
\end{figure}

\begin{figure}[tbp]
	\centering
	\includegraphics[width=\textwidth]{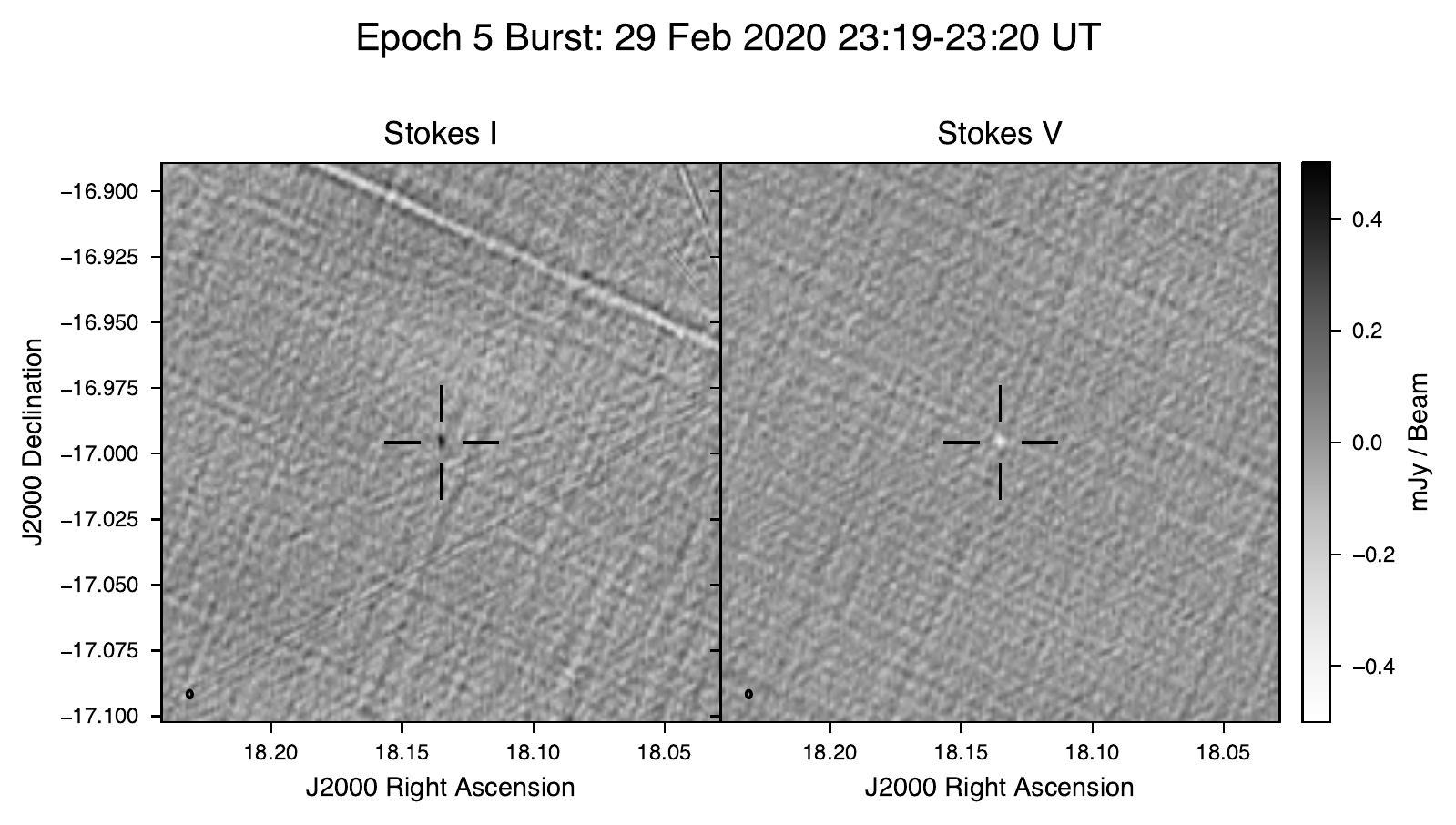}
	\caption{2-4~GHz image of YZ~Cet in Stokes~I \textit{(left)} and Stokes~V \textit{(right)} during the left-polarized coherent burst in Epoch~5, following the approach of Figure~\ref{fig:epoch2burst_image}. We measure an average flux density of 468$\pm$70~$\mu$Jy (I) and -441$\pm$55~$\mu$Jy (V) for YZ~Cet during this interval. \label{fig:epoch5burst_image}}
\end{figure}

After imaging, we masked the star out of the model so that the model contained only background sources, then subtracted the model from the visibility data to obtain residual visibilities containing only the star and noise. Figures~\ref{fig:epoch2burst_image} and~\ref{fig:epoch5burst_image} show images of radio bursts in Epochs 2 and 5 in the background-subtracted data.  In Stokes~I, residual sidelobes are visible due to imperfect subtraction of PMN~J0112-1658.

\begin{figure*}[tbp]
	\centering
	\includegraphics[width=0.8\textwidth]{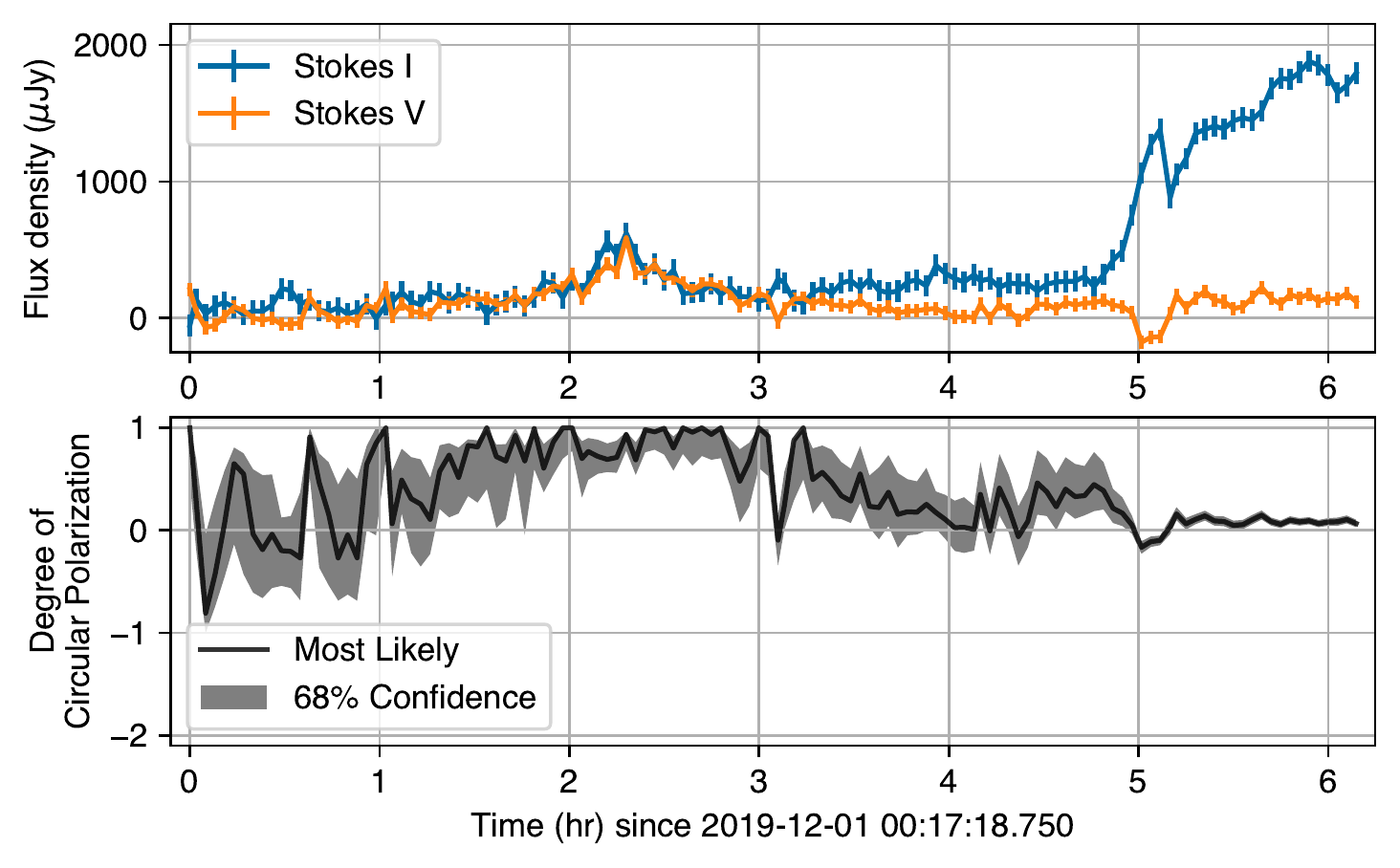}
	\caption{2-4~GHz time series of YZ~Cet during Epoch 2 (1 Dec 2019), with 3-minute integrations. \textit{(Top)} Stokes~I and V flux density. \textit{(Bottom)} Degree of circular polarization, showing the highest likelihood value (black) and 68\% confidence interval (gray), which was derived through a maximum likelihood approach described in the text.  The coherent burst at 2.3 hours has right circular polarization of $93.5^{+2.6}_{-15.9}$\% in the time bin with peak flux density, as shown on the plot.  The large event at 5-6 hours is likely an incoherent gyrosynchrotron flare based on its low polarization. A smaller event at 5.1~hours is also a likely coherent burst,
    whose strong left circular polarization causes a temporary reversal of the total polarization. \label{fig:epoch2_tseries}}
\end{figure*}

\begin{figure*}[tbp]
	\centering
	\includegraphics[width=0.8\textwidth]{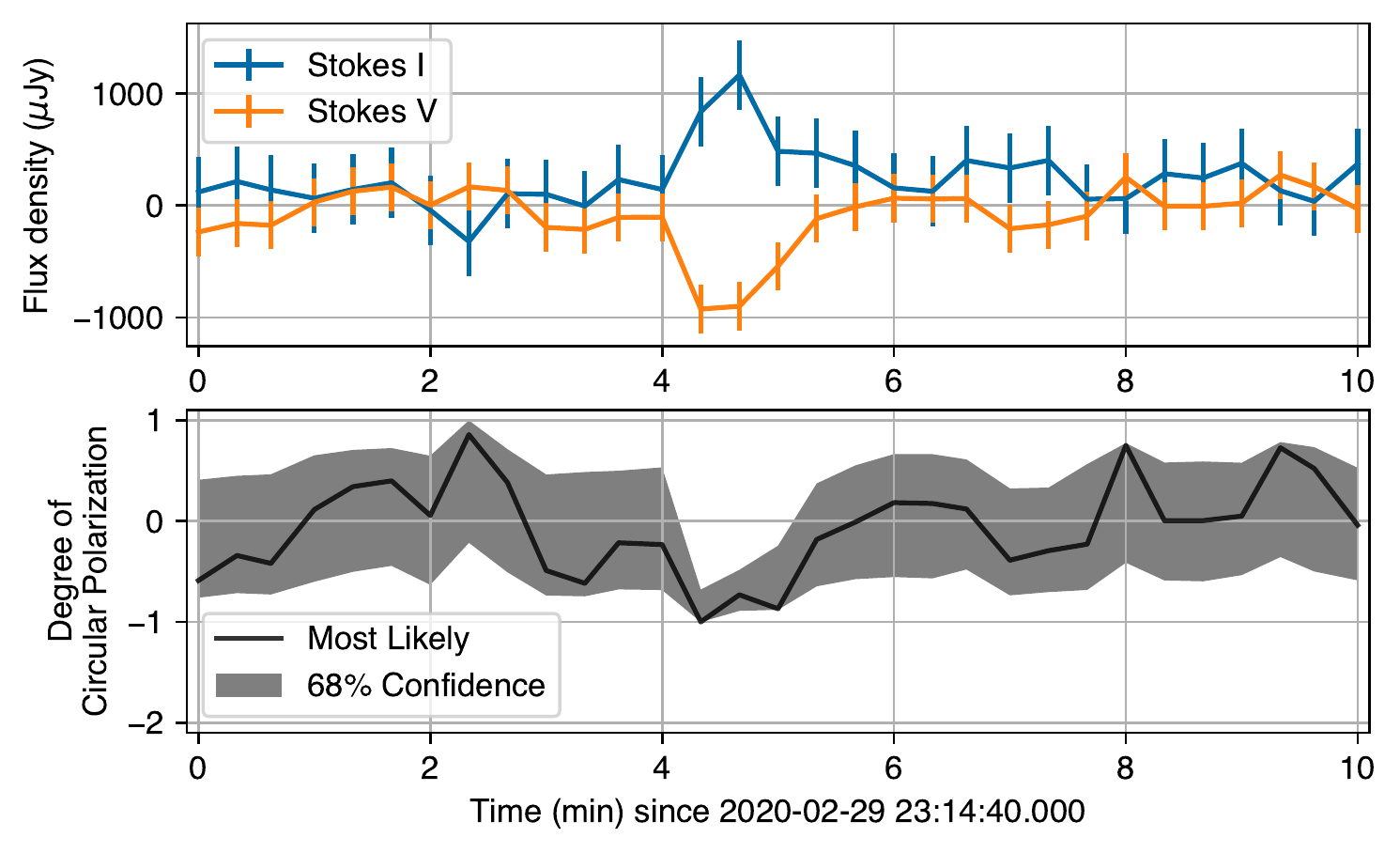}
	\caption{2-4~GHz time series of YZ~Cet during part of Epoch 5 (29 Feb 2020), with 20-second integrations. \textit{(top)} Stokes~I and V flux density. \textit{(bottom)} Degree of circular polarization, showing the highest likelihood value (black) and 68\% confidence interval (gray), which was derived through a maximum likelihood approach described in the text. The coherent burst has left circular polarization of $73^{+16}_{-25}$\% in the time bin with peak flux density (4.7~min), as shown on the plot. \label{fig:epoch5_tseries}}
\end{figure*}

With the star at the phase center, we used the \textbf{plotms} task to average the residual visibilities across all baselines and frequencies, yielding a complex-valued time series.  The real component is equivalent to the center pixel of a natural-weighted image, yielding the flux density of the star.  The imaginary component should not contain stellar flux, but exhibits comparable levels of noise due to thermal noise, RFI, and sidelobes of imperfectly-subtracted background sources.  These residual background sidelobes can cause ``ripples'' in the time series as the sidelobe pattern evolves over time.  We calculated the standard deviation of the imaginary component to estimate the effective noise levels in the time series including these factors.
We used the imaginary component of the time series to estimate noise levels because, in epochs without clearly detected stellar variability, the real and imaginary components have roughly similar standard deviations. For example, in Epoch 1, the Stokes I standard deviation is 69~$\mu$Jy (real) and 79~$\mu$Jy (imaginary), and the Stokes~V standard deviation is 50.5~$\mu$Jy (both real and imaginary).  The greater standard deviation for Stokes~I than V illustrates the effect of residual sidelobes of background sources.

For three-minute integrations, we measured noise levels of 55-80~$\mu$Jy in the Stokes~I time series and 37-50$\mu$Jy in Stokes~V.  Without source confusion, the VLA's theoretical sensitivity in three minutes is 22~$\mu$Jy. Since source confusion is not an issue in Stokes~V, the elevated Stokes~V noise levels are likely due to RFI, both data loss due to RFI flagging and low-level RFI in the remaining data.  The Stokes~I noise levels are affected by both RFI and imperfect background source subtraction; both of these effects are enhanced by the VLA's compact configuration.

Figure~\ref{fig:phased_tseries} shows the resulting time series as a function of orbital phase, where the shaded region shows $\pm$3 times the estimated noise level on flux density in each of the 3-minute time bins in that epoch. Figures~\ref{fig:epoch2_tseries} and~\ref{fig:epoch5_tseries} show detailed time series of Epochs 2 and 5, the two epochs with coherent bursts. To identify bursts, we required a flux increase of $>3\sigma$ during the burst compared to before or after the burst. For example, in Epoch~2 (Figure~\ref{fig:epoch2_tseries}), the events at 2.3, 5.1, and 5-6 hours satisfy this criterion, whereas a possible left-polarized event at 3.1~hours constitutes only a $2\sigma$ flux enhancement.

For epochs without bursts, we measured or placed an upper limit on the quiescent emission levels using an image of the full epoch duration after background source subtraction.  For Epoch~1, we measured an intensity of $I=-39$~$\mu$Jy/beam in the image at the star's location (star undetected) and an RMS in the image near the star's location of $\sigma$=25~$\mu$Jy/beam, leading to a 3$\sigma$ upper limit on source flux density of: $S< 3\sigma=75$~$\mu$Jy.  In Epoch~3, the star was detected with a peak flux density of $313\pm20$~$\mu$Jy (Stokes I) and $18.7\pm5.3$~$\mu$Jy (V).  In Epoch~4, the star was undetected with an intensity of $36\pm21$~$\mu$Jy at its location in the image, yielding a 3$\sigma$ upper limit of 64~$\mu$Jy on flux density. Weakly polarized, non-thermal, slowly varying quiescent emission on M~dwarfs, such as in Epoch~3, is typically attributed to incoherent gyrosynchrotron emission \citep{Benz1994A&A...285..621B}.

To assess the bursts' degree of circular polarization, $r_c = V/I = (RR-LL)/(RR+LL)$, we used a maximum likelihood approach to estimate $r_c$ and construct a 68\% confidence interval. To calculate the likelihood, we assumed RR and LL are Gaussian-distributed, obtaining the standard deviation for each from the imaginary component of the time series.  We generated a probability distribution function (PDF) for obtaining the data in terms of model parameters $S_I$ (Stokes~I flux density) and $r_c$, then marginalized the distribution across $S_I$ to obtain a PDF for $r_c$ alone. The black line in Figures~\ref{fig:epoch2_tseries} and~\ref{fig:epoch5_tseries} shows the value of $r_c$ at which the PDF peaks, and the gray confidence interval shows the range of $r_c$ that lie from 0.16 to 0.84 in the cumulative distribution function.

\begin{figure}[tbp]
	\centering
    \includegraphics[width=\textwidth]{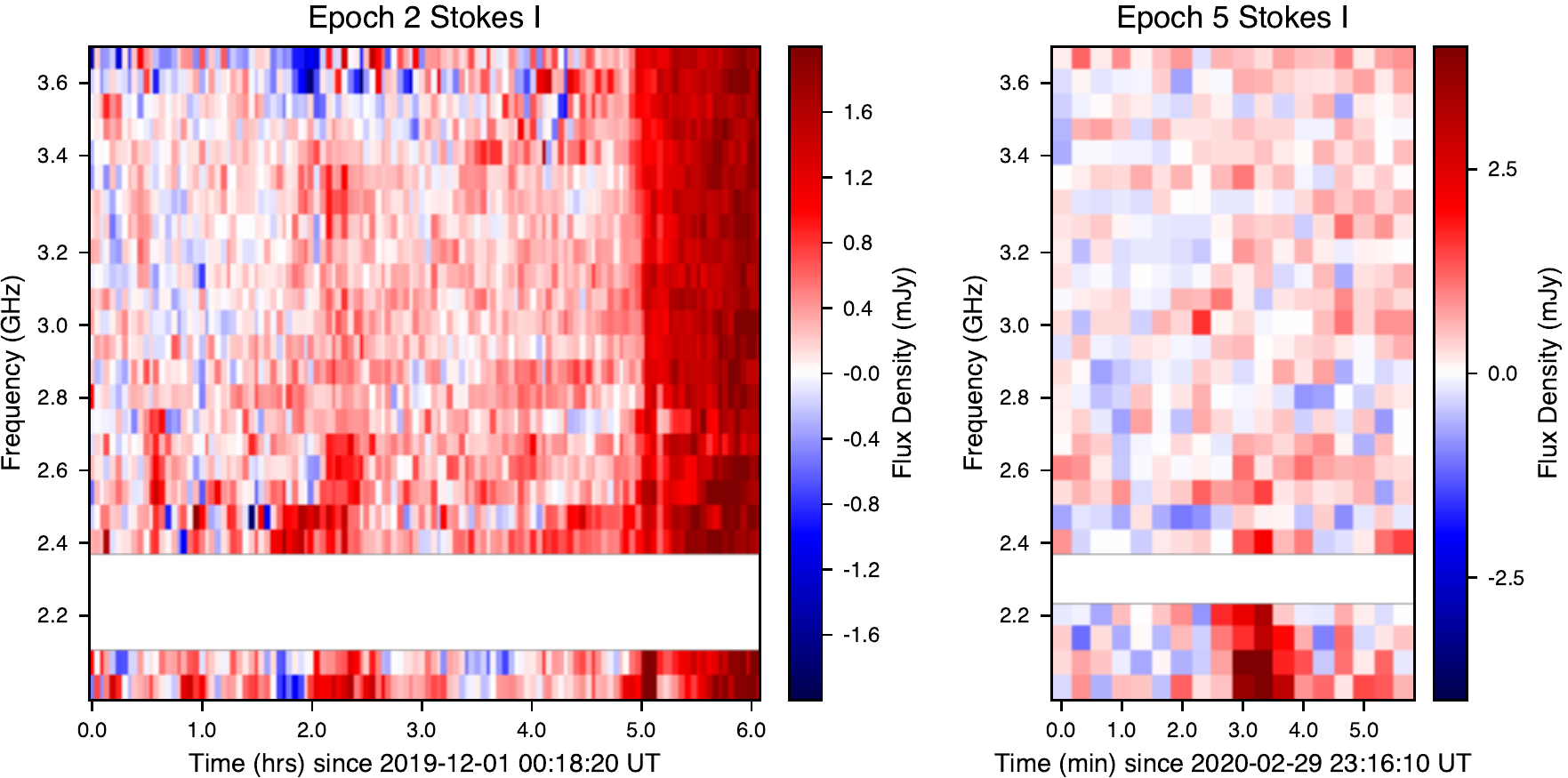}
	\caption{2-3.75~GHz Stokes I dynamic spectra of bursting epochs.  \textit{(Left)} All 6.5 hours of Epoch~2, with an uncertainty of 265~$\mu$Jy in each pixel of 3 minutes and 64 MHz. The broadband radio flare dominates the spectrum near the end of Epoch~2. Above $\sim$3.4~GHz the noise increases due to RFI. In Stokes~I, residual sidelobes of background sources create artificial structure, such as diagonal striping and negative (blue) flux. \textit{(Right)} 6-minute excerpt of Epoch~5, with an uncertainty of 620~$\mu$Jy in each pixel of 20 sec and 64 MHz.\label{fig:dynI}}
\end{figure}

We produced dynamic spectra (Figure~\ref{fig:dyn} and Figure~\ref{fig:dynI}) for all of Epoch 2 and for a short time period surrounding the Epoch 5 burst, using baseline-averaging code described in \cite{Villadsen2019}. The flux density uncertainties quoted in the caption are calculated using the imaginary component of the dynamic spectrum (which does not contain stellar emission), by taking the standard deviation in each frequency channel then calculating the median across all channels. The peak flux densities in Stokes~V for the Epoch 2 bursts at 2.3 hours and 5.1 hours both exceed 5$\sigma$, as does the Epoch 5 burst. The incoherent flare in Epoch 2 has weak right polarization, appearing only faintly in Stokes~V except for the coincident LCP burst at 5.1 hours, and in Stokes I it spans the entire 2-3.7~GHz band, consistent with the broadband nature of gyrosynchrotron emission.

In Epoch 2, the right-polarized burst and the left-polarized burst at 5.1 hours both are brightest at lowest frequencies ($<$3~GHz). Like the two clearest features in the Epoch 2 dynamic spectrum, the Epoch 5 event is also brightest at the lowest frequencies.  These three events that appear clearly in the dynamic spectra drop off above 2.5-3~GHz.  If the emission process is the cyclotron maser, this indicates that the maximum magnetic field in the source regions is of order 1~kG.

% solar mass-loss : 2e-14 msun/year

\begin{figure*}[tbp]
	\centering
	\includegraphics[width=0.5\textwidth]{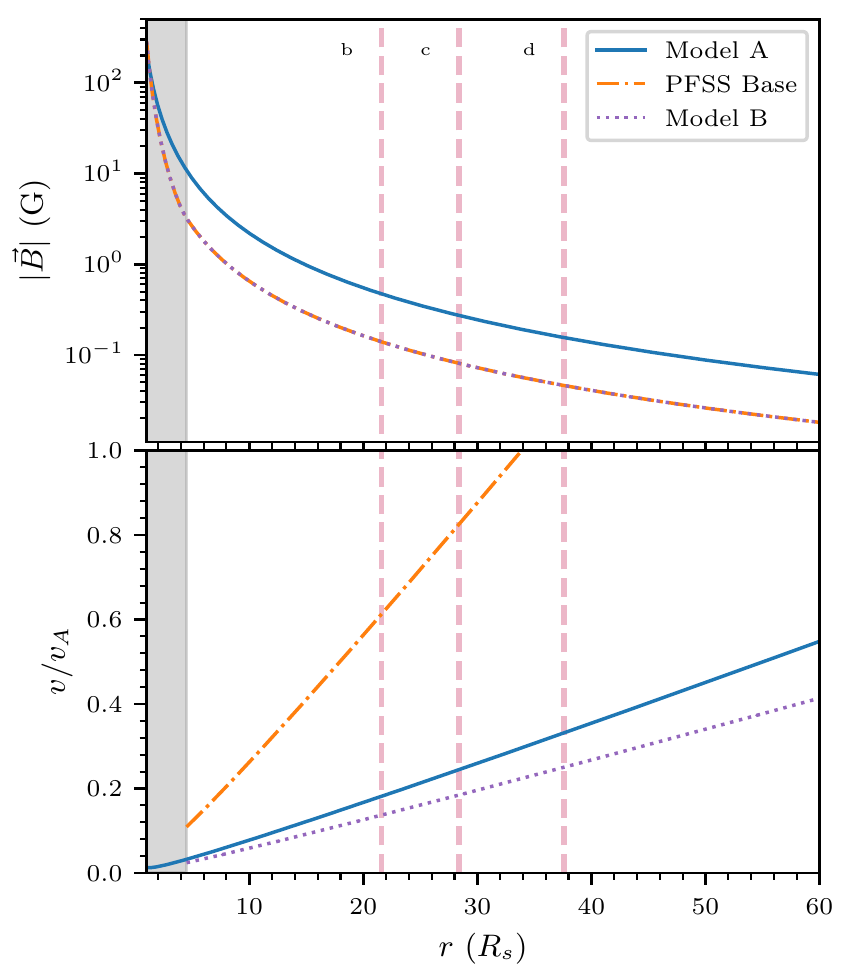} % requires the graphicx package
	\caption{Magnetic properties of the circumstellar environment in our two fiducial models. \textit{Top} - The magnetic field strengths across the YZ~Ceti planetary environment are different between our fiducial models, A: a radial field declining from the surface with strong mass-loss (solid line), and B: a magnetically weaker potential field source surface (PFSS) extrapolation with weaker mass-loss rate (dotted line). For comparison, the PFSS base model (dash-dot line), shows the intermediate case of a weaker PFSS model magnetic field, utilizing the higher mass-loss rate of model A. \textit{Bottom} - The Alfv\'{e}nic Mach number (relative speed in planet frame assuming circular Keplerian orbits), reveals the predominant interaction regime for our fiducial models as a function of distance from host (in units of stellar radii). The gray region at left denotes the closed field region used for the PFSS in model~B. The magnetic mach number at the YZ~Ceti planets (locations shown as vertical dashed lines) is sensitive to the combination of assumed magnetic field strength and stellar mass-loss rate, with each planet in the system firmly within the sub-Alfv\'{e}nic regime in both models A and B.}
	\label{fig:fieldcomp}
\end{figure*}

\begin{figure*}[tbp]
	\centering
	\includegraphics[width=0.5\textwidth]{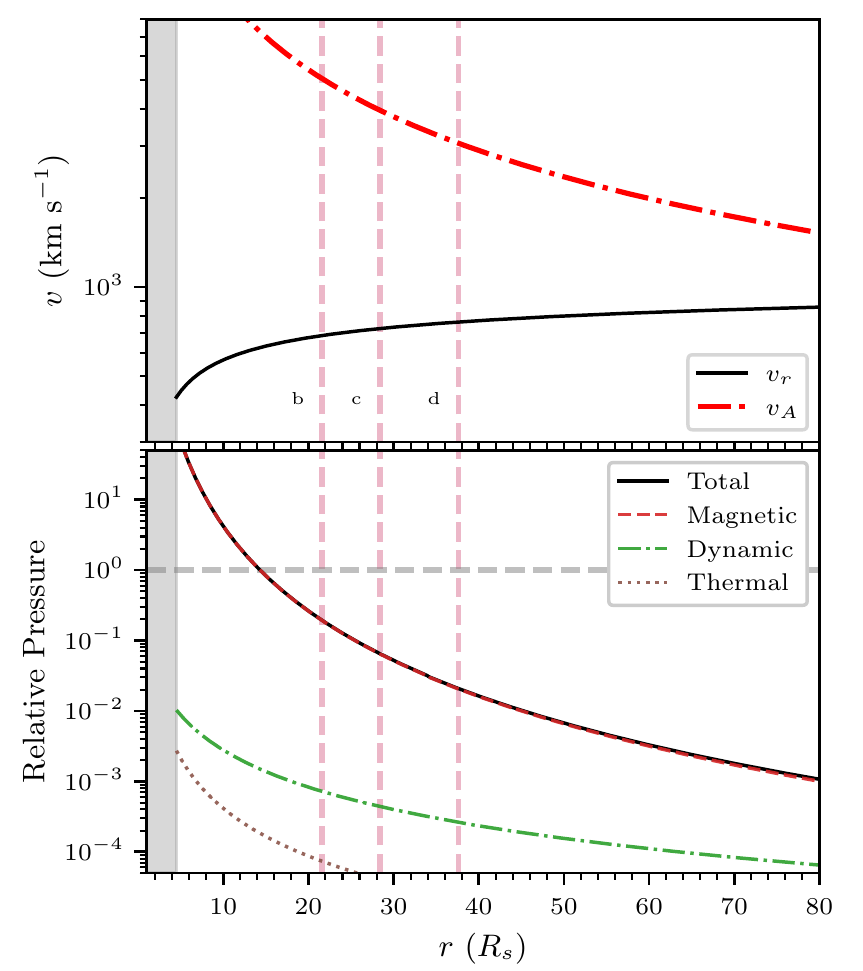} % requires the graphicx package
	\caption{Stellar wind properties in our two fiducial models. Our main equatorial isothermal wind model for the YZ~Ceti environment (Model B) shows that the planetary system (located at distances corresponding to vertical dashed lines) is likely to be interacting in the sub-Alfv\'{e}nic regime with the host star. The radial wind velocity is 600-700 km s$^{-1}$ near the location of the planets (see \textit{Top}), and well below the corresponding Alfv\'{e}n speed. At the location of the planets the total pressure from the assumed stellar wind (see \textit{Bottom}), plotted relative to the magnetic pressure of a 1~G field, is below that reference pressure (horizontal dashed line). Consequently, appreciable planetary magnetic fields on these planets ($\gtrsim$ 0.3~G) are likely to carve out their own planetary magnetospheres. The gray region at left denotes the closed field region used for the PFSS.}
	\label{fig:windmodel}
\end{figure*}

\noindent \textbf{Stellar Magnetospheric Environment.} In order to determine whether star-planet interactions could have powered our observed polarized radio emissions we needed to characterize the likely magnetospheric environment impacting the YZ~Ceti planetary system. We considered two models: A) a magnetosphere defined by a radial isothermal stellar wind whose properties are set by the corona and the surface magnetic field strength, and B) a potential field source surface (PFSS) extrapolation of typical M-dwarf Zeeman Doppler Imaging measurements including an isothermal stellar wind solution beyond the source surface \citep{See2017}. Since the magnetic field and wind environments of low-mass stars are very uncertain, this approach explores the effects of a range of likely stellar magnetic field strengths experienced by the YZ~Ceti planets. 

The first approach, often employed in the literature, uses a stellar wind originating from the stellar surface, which overestimates the magnetic field at the planet location because it does not take into account the rapid radial decay of closed field lines near the stellar surface \citep[e.g.,][]{Turnpenney2018}. The second approach accounts for this effect by using a more realistic stellar magnetic field topology \citep[e.g.,][]{See2017}; however, the inherent assumptions exclude additional stresses to the magnetic field, and may underestimate the strength of the magnetic field at planetary distances from the star, beyond a poorly constrained source surface.

\textit{Model A: Radial Stellar Field, High $\dot{M}$.} To formulate the radial stellar wind solution we used a Weber wind model \cite{Weber1967}, which solves the ideal magnetohydrodynamics problem in spherical coordinates for an axisymmetric equatorial wind propelled by a rotating star. Note that there is a typo in their equation 23. In the last term of the denominator in brackets, the factor $\Omega^{2} r^{2} M_{A}^2$ should be $\Omega^{2} r^{2} M_{A}^4$. Their solution is similar to that of \cite{Parker1958}, but self-consistently incorporates the stresses of the ionized wind on the magnetic field anchored to the rotating star. We depart from \cite{Weber1967}, by employing an isothermal wind, a sub-case of their general polytropic approach. The physical wind solution is the one which smoothly passes through the three critical points (1 sonic, 2 magnetic) \citep{Weber1967}, constraining the solution and fixing the initial radial wind velocity for our choice of boundary conditions. The coupled solution requires as inputs: stellar mass, radius, and rotation rate (Table~\ref{tab:prop}), as well as coronal plasma temperature, an average mass-loss rate, and the radial magnetic field strength at the surface (see below). With these assumptions, we numerically solved for the wind radial profile. In practice, the system of equations incorporating the critical points fixes the total energy, a constant of the motion. We then employed the energy equation to numerically solve for the radial speed as a function of distance from the star. The other system properties could then be determined from the radial wind profile \citep{Weber1967}. In summary, the significant variable parameters for the boundary conditions reduce to the coronal temperature, radial magnetic field strength, and mass-loss rate.

For this model (A) we assumed a coronal temperature of $kT = 0.25$ keV $\approx 3 \times 10^{6}$ K, constant mass-loss rate of  $\dot{M} \equiv 4\pi \rho u {r}^{2} = 10^{-13}$ $M_{\odot}$ yr$^{-1}$  $=  5 \, \dot{M}_{\odot}$ (fives times the mass-loss rate of the Sun), and radial field of $B_{r} =$ 220~G. The coronal temperature is similar to that of other inactive late M-dwarfs, based on their X-ray observations \citep[e.g.,][]{Stelzer2013,Loyd2016ApJ...824..102L}. The mass-loss rate is a compromise between the expected rates for similar stars based on their Rossby numbers ($Ro = 0.5$ for YZ~Ceti) \citep{See2017}, and the low rate of Prox Cen \citep[see within][]{Kavanagh2021MNRAS.504.1511K}, which has similar physical properties, albeit with a weaker magnetic field strength (see Table~\ref{tab:prop}).

For the estimate of the average radial surface magnetic field strength of YZ~Ceti, we utilized the measured large-scale field topology of Prox Cen, from \cite{Klein2021MNRAS.500.1844K}, since no such measurements of YZ~Ceti are yet available. Because of their similar properties (Table~\ref{tab:prop}), Prox Cen is a useful analog for interpreting the magnetic properties of YZ~Ceti, and is one of the few slowly rotating late M-dwarfs with a measured field topology from Zeeman Doppler Imaging. We scaled the measured magnetic field of Prox Cen, in its spherical harmonic decomposition, based on the measured Zeeman Broadening measurement of YZ~Ceti, $<B_{\mathrm{ZB}}> = 2200$~G \citep{Moutou2017}.
We defined the scaling to achieve an average field flux strength ratio of $\zeta \equiv \, <B_{\mathrm{ ZDI} }> / < B_{ \mathrm{ZB} }> \, \approx \, 0.1$, and thus an average radial surface field of 220~G. We chose $\zeta=0.1$ as a representative value for the sample of stars  with similar properties in \cite{Morin2010MNRAS.407.2269M} that have both kinds of Zeeman measurements. $\zeta$'s low value originates from field cancellation in the Stokes V ZDI measurements, as opposed to the Stokes I ZB measurements that include the total field strength. It is worth noting that YZ~Ceti's Zeeman broadening measurement is a high outlier for its Rossby number $\sim$0.5 \citep{Reiners2022A&A...662A..41R}, and the source measurements \cite{Moutou2017} may be systematically high \citep{Reiners2022A&A...662A..41R}, especially for slow rotators.  However, ZDI has measured M~dwarf $\zeta$ values up to $\sim$0.3 \citep[Prox Cen][]{Klein2021MNRAS.500.1844K}, so our estimated average large-scale field of 220~G may be reasonable even if the current YZ~Ceti ZB measurement is an overestimate. With an average large-scale field of 220~G, surface variations and small-scale fields in the low stellar corona could still lead to regions with kG field strengths, plausibly allowing ECM emission at 2-3~GHz.

\textit{Model B: PFSS Stellar Field, Low $\dot{M}$.} For model B we used the same assumptions for the wind properties with two key differences. The first is that we moved the inner boundary of the Weber wind model, where the stellar field is purely radial, to 4.5 stellar radii, consistent with MHD simulations of M dwarf winds \citep[Table 2 in][]{Vidotto2014MNRAS.438.1162V}.  Shifting this ``source surface'' outwards accounts for closed magnetic field lines near the surface. We modeled this closed field by filling the space between the stellar surface and the wind source surface with a PFSS extrapolation \citep[e.g.,][]{Altschuler1969SoPh....9..131A} based on Prox.\ Cen.'s field topology, scaled to yield an average large-scale radial field of 220~G at the stellar surface. The PFSS extrapolation thus sets the average radial magnetic field strength at 4.5 stellar radii from the star. Secondly, we also assumed a 20-times lower mass-loss rate of $0.25$ $\dot{M}_{\odot}$, comparable to the upper limit on Prox Cen's wind from \cite{Wood2001ApJ...547L..49W}, and consistent with the predicted rate from \cite{Kavanagh2021MNRAS.504.1511K}. This value is also close to the expectation ($\sim$0.23~$\dot{M}_{\odot}$) calculated from the relation between X-ray surface flux and mass-loss rate \citep{Wood2021ApJ...915...37W,Vidotto2021LRSP...18....3V}. We compare these distinct magnetic environments in Figure~\ref{fig:fieldcomp}.

We further illustrate some properties of model B in Figure~\ref{fig:windmodel}, showing the significant velocities relevant to the wind (\textit{Top} panel), and the total wind pressure throughout the model environment around YZ~Ceti. We considered model B to correspond to our most realistic estimate of the average magnetized environment pervading this planetary system, whereas model A encapsulates typical assumptions in the literature treatment of these questions. While informed by the literature, the wind parameters are typically uncertain for low-mass stars, but since we employed an analytic model, we can readily change the input assumptions to determine their effect on the potential for the YZ~Cet planetary system to power radio emissions (see below). To provide some intuition for the isothermal wind solution and the impacts of these parameter assumptions, we note that changing the temperature is the most significant parameter determining the wind velocity, changes in the mass-loss rate largely impact the wind density, and the radial field strength scales the overall magnetic field since the azimuthal field component is much weaker for slowly rotating systems. In the absence of a 3D wind simulation \citep[e.g.,][]{Kavanagh2021MNRAS.504.1511K}, these simplified isothermal approaches provide a reasonable means to examine the approximate interplanetary environment conditions \citep{Vidotto2021LRSP...18....3V}.

\noindent \textbf{Planet Induced Radio Emission.} Our detection of polarized radio bursts from YZ~Ceti prompts the question of whether the coherent radio emission could have been powered by the magnetic interaction of the star with its planets \citep[see within][]{Zarka2007}. We used Models A and B (described above), to define the magnetized stellar wind filling the environment of the YZ~Ceti planetary system. When this wind interacts with the planets, the dissipated energy can power auroral radio emissions. We estimated the available power through this interaction using the frameworks of \citep{Lanza2009} (Reconnection), and \citep{Saur2013} (Alfv\'en Wings), similar to the approach taken by \citep{Vedantham2020}.

The available power released by magnetic reconnection \citep{Lanza2009} is 

\begin{equation}
    S_{l} = \frac{1}{4} \gamma R_{o}^{2} \upsilon B^{2} \; ,
    \label{eq:lanza}
\end{equation}

\noindent in cgs units, where $\gamma$ is a geometric factor, $R_{o}$ is the radius of the obstacle, i.e., the planetary magnetosphere, $\upsilon$ is the velocity of interaction in the frame of the planet, and $B$ is the star's magnetic field strength at the planet location. Similarly, the available power transmitted through Alfv\'en wings \cite{Saur2013}, a prediction valid in the low Mach number regime, is

\begin{equation}
    S_{s} = \frac{1}{2} \bar{\alpha}^{2} R_{o}^{2} \upsilon^{2} B \sin^{2} \theta \sqrt{4\pi \rho} \; ,
    \label{eq:saur}
\end{equation}

\noindent as expressed by \citep{Pineda2018}, where $\bar{\alpha}$ is an interaction strength, $\theta$ indicates the angle between the wind's relative velocity vector and the magnetic field in the planet's frame of reference, and $\rho$ is the mass density of the magnetized flow. These two approaches differ by a factor of twice the Alfv\'{e}nic Mach number \citep{Saur2013} and a geometric factor. We take the wind properties for models A and B, and use these expressions to estimate the expected power available to generate ECM radio emissions. At the planet's location, the wind and magnetic field are aligned and nearly radial due to the star's slow rotation, but the planet's orbital velocity (small relative to wind speed) gives $\theta$ a small, non-zero value. We focused on YZ~Ceti~b as the closest-in planet and most likely to power the detected hundreds of $\mu$Jy bursts in our radio data sets.

In evaluating Equations~\ref{eq:lanza} and~\ref{eq:saur}, we take the middle value of the geometric factor, so $\gamma \rightarrow 1/2$, and consider the interaction strength $\bar{\alpha} \rightarrow 1$. The former is justified by our ignorance of the exact geometry of the interacting magnetic fields \citep{Lanza2009}. For the latter, we justify the assumption of the interaction strength based the likely conductivity of the planetary obstacle through its magnetosphere or ionosphere, considering the environments of the large close-in rocky planets of the YZ~Ceti system to have high Pederson conductivities, see Appendix A of \cite{Turnpenney2018}.

The last remaining variable in the power expressions is the planetary obstacle radius, $R_{o}$. This is defined by the size of the planetary magnetosphere, or at a minimum the radius of the planet itself assuming a thin ionosphere. We use the pressure balance between the supposed planetary field and the wind to define the radius of the planetary magnetopause:

\begin{equation}
    R_{o} = R_{p} \left(  \frac{  B_{p}^{2}  }{ 8\pi \rho \upsilon^{2}  + 8\pi \rho k T/\mu m_p  + B^{2}  }      \right)^{1/6} \; ,
    \label{eq:ro}
\end{equation}

\noindent where $B_{p}$ is the assumed planetary dipole field strength, $\mu = 0.5$ for a fully ionized hydrogen wind, and $m_{p}$ is the proton mass. If the ratio of $R_{o} / R_{p}$ from Equation~\ref{eq:ro} falls below unity, we instead use $R_{p}$ as the obstacle radius. 

The YZ~Ceti system was characterized with radial velocity measurements and does not exhibit transits, so the planet radii are unknown. The planets are likely to be roughly Earth-sized, and YZ~Ceti~b has a minimum mass of 0.7~$M_{\oplus}$. For the radii of YZ~Ceti~b we consider a range from $R_{p} = 0.89$~$R_{\oplus}$ to $R_{p} = 1$ $R_{\oplus}$, where the lower bound corresponds to the radius of the minimum mass assuming it also has an Earth-like density. As the planet is roughly Earth sized we explore a range of planetary dipole field strengths starting from 1~G (Earth-like), increasing it by an order of magnitude (10~G), and decreasing it to below the stellar field strength at the planet location (effectively unmagnetized). These values set the abscissa range in Figure~\ref{fig:fluxden}.

With these assumptions, we can compute the energy available to power auroral radio bursts from YZ~Ceti, using both the Reconnection and Alfv\'{e}n Wing prescriptions, as well as considering both the Model A and Model B wind environments. To convert the power to a possible burst radio flux density we use
\begin{equation}
    F_{\nu} = \frac{ \epsilon S}{\Omega \Delta \nu d^{2}} \; ,
\end{equation}
\noindent where $S$ comes from Equation~\ref{eq:lanza} or~\ref{eq:saur}, $\epsilon = 0.01$ is the radio efficiency factor \citep{Zarka2007,Saur2013}, $\Delta \nu = 3$~GHz is the emission bandwith, for which we assume the emission spans from low frequencies to our emission band, $d = 3.712$~pc is the distance to the star, and we use $\Omega = 0.16$~sr for the beaming angle based on the observed value for Jupiter-Io radio emission \citep{Queinnec2001P&SS...49..365Q}. The results of our calculations are displayed in Figure~\ref{fig:fluxden}, and discussed in the main text.

\begin{figure*}[tbp]
	\centering
	\includegraphics[width=0.8\textwidth]{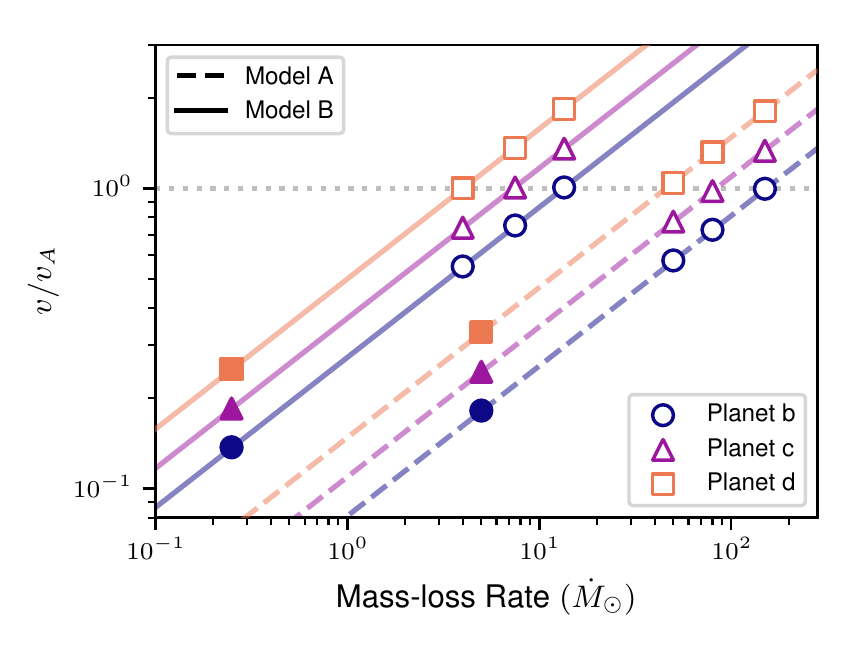} % requires the graphicx package
	\caption{Alfv\'en Mach number for YZ Ceti's planets as a function of stellar wind mass loss rate.  Starting from our fiducial Models A and B, the Alfv\'{e}nic Mach number at the location of each planet will grow with the assumed mass-loss rate in the isothermal wind calculations (shown in units of the mass-loss rate of the Sun, $2\times 10^{-14} M_{\odot}$ yr$^{-1}$). The set of points for select mass-loss rates give the Alfv\'{e}nic Mach number at the location of each of the planets YZ Ceti bcd, with the filled symbols denoting our fiducial models. The lines follow the approximation $\upsilon/ \upsilon_{A} \approx \sqrt{\upsilon_{r} \dot{M}} / B_{r} $, where the velocity and field conditions are evaluated at the location of each planet. Tracing the lines for each planet (using Model A or B), to $\upsilon/ \upsilon_{A} =1$ (dotted-line) gives the mass-loss rate at which each planet passes beyond the Alfv\'{e}n surface.}
	\label{fig:massloss}
\end{figure*}

\noindent \textbf{SPI Parameter Space.}  The predicted flux densities for star-planet interaction depend on a variety of unknown properties for the magnetized environment, most prominently the assumed stellar mass-loss rate, and the stellar field strength.
%We have undertaken an approach to define values consistent with the literature and the known physical properties of the star.
Above, we chose Models A and B to represent a range of values consistent with the literature and the known physical properties of the star. Below, we explore two specific effects related to these assumptions: the range of mass-loss rates consistent with the SPI scenario and the dependence of the SPI power on the assumed stellar magnetic field.

\textit{Constraints on mass-loss rate:} If our detected bursts are indeed powered by sub-Alfv\'{e}nic star-planet interaction, then the corresponding planet must be within the Alfv\'{e}n surface of the stellar environment. Using our isothermal wind solution, we explored the impact of the assumed mass-loss rate on the Alfv\'{e}nic Mach number at the position of the planets around YZ~Ceti. We show these results in Figure~\ref{fig:massloss}. Both fiducial Models A and B allow for an increase of an order of magnitude in the assumed mass-loss rate before any planets go beyond the Alfv\'{e}n surface, and even more before YZ~Ceti~b becomes super Alfv\'{e}nic.

Because the star is rotating slowly, the wind speed and stellar field are largely radial, and since the orbital motion of the planets is small in comparison to the wind speed, the results in Figure~\ref{fig:massloss} are well approximated by
\begin{equation}
    \frac{\upsilon}{\upsilon_{A}} \approx \frac{ \sqrt{\dot{M} \upsilon_{r}} }{ B_{r}} \; ,
    \label{eq:mach}
\end{equation}
\noindent where the velocity and field on the right-hand side correspond to the radial components evaluated at the position of the planets. For our fiducial Model~A (5$\dot{M}_{\odot}$) the planets a, b, and c, become super Alfv\'{e}nic at mass-loss rates of approximately 150, 80, and 50 times $\dot{M}_{\odot}$, respectively. For our fiducial Model~B (0.25$\dot{M}_{\odot}$) the planets a, b, and c, become super Alfv\'{e}nic at mass-loss rates of approximately 13.5, 7.5, and 4 times $\dot{M}_{\odot}$, respectively. 
Increasing the assumed stellar magnetic field would increase the distance corresponding to the Alfv\'{e}n surface, and increase the planets' corresponding sub/super-Alfv\'{e}nic transition mass-loss rate. If our radio detections are powered by the interaction of YZ~Ceti~b with its host, it should imply that YZ~Ceti has a mass-loss rate within these bounds, likely $<13.5$ $\dot{M}_{\odot}$, using our more realistic Model~B stellar field topology.

\textit{Scaling the stellar magnetic field:} As discussed when introducing Models~A and~B above, we conservatively assumed a surface radial average field strength of 220~G, but this may be an underestimate. For both our Models A and B, scaling the magnetic field at the surface toward higher values linearly scales the field at the location of the planet, and similarly scales the minimum predicted SPI flux density as a function of planet field strength (flat regions in Figure~\ref{fig:fluxden}). The turning points in Figure~\ref{fig:fluxden} also shift toward higher planet field strengths, as their position encodes where the stellar and planetary fields balance. 

For Model~A, a stellar field increase of a factor of 2 or 3 places flux density predictions above the measured bursts, for both the Reconnection and Alfv\'en Wing mechanisms. If our radio detections are indeed SPI, this underscores
that Model~A overestimates the stellar field strength at the planet's location by not accounting for closed field structures.
%to which the assumed Model~A environment is already overestimating the stellar field strength at the position of the planet.
Literature SPI predictions with stellar radial fields, like what we have assumed with Model~A, may over-predict both SPI intensities and the size of the Alfv\'{e}n surface and which planets it encompasses.

For Model~B, an increase in the field strength of a factor of 2 or 3 pushes both the Reconnection and Alfv\'{e}n Wing predictions to higher values. While the Reconnection prediction would become more discrepant with the measured burst flux densities, it would place the Alfv\'{e}n Wing prediction consistent with smaller (but non-negligible) planetary dipole fields.
If our detections are indeed SPI, and if additional measurements reveal that YZ~Cet's average global field significantly exceeds 220~G, then our result would support Model~B (accounting for closed field near the surface) over Model~A (open stellar field), and the Alfv\'en Wing mechanism over the Reconnection mechanism.
%If our detections are indeed SPI, and additional measurements of the stellar surface magnetic field reveal our assumption as too conservative, our result highlights that the combination of employing Alfv\'{e}n Wing predictions with a stellar field that accounts for closed-field near the surface, represents the most accurate means of assessing the power available through star-planet interactions. 

%If our detections are indeed SPI, and if additional measurements reveal that YZ~Cet's average global field significantly exceeds 220~G, then our result highlights that the combination of employing Alfv\'{e}n Wing predictions with a stellar field that accounts for closed-field near the surface represents the most accurate means of assessing the power available through star-planet interactions. 

\section*{Data Availability}

The radio data used in this publication are available through the NRAO archive (\url{data.nrao.edu}) under project code VLA/19B-222.

\section*{Code Availability}

The raw radio data were processed with publicly available software package \texttt{CASA} \citep{mcmullin2007} and NRAO's VLA calibration pipeline. The codes describing the model stellar wind implementation are available upon reasonable request to J.S.P. The public python packages that are part of \texttt{astropy} also aided in the analysis and presentation of results \citep{astropy2013A&A...558A..33A,astropy2018AJ....156..123A}.

\section*{Acknowledgements}

%Acknowledgements should be brief, and should not include thanks to anonymous referees and editors, or effusive comments. Grant or contribution numbers may be acknowledged.

The authors thank Baptiste Klein and Julien Morin for providing spherical harmonic coefficients for Prox.\ Cen's field topology. We also thank Aline Vidotto, Joachim Saur, Rim Fares, Moira Jardine, and Antoaneta Antonova, for useful discussions in the preparation of this article.

This material is based upon work supported by the National Science Foundation under Grant No.\ AST-2108985 (JSP) and AST-2150703 (JRV). The National Radio Astronomy Observatory is a facility of the National Science Foundation operated under cooperative agreement by Associated Universities, Inc. This paper includes data collected by the TESS mission. Funding for the TESS mission is provided by the NASA's Science Mission Directorate. This work has made use of data from the European Space Agency (ESA) mission
{\it Gaia} (\url{https://www.cosmos.esa.int/gaia}), processed by the {\it Gaia}
Data Processing and Analysis Consortium (DPAC,
\url{https://www.cosmos.esa.int/web/gaia/dpac/consortium}). Funding for the DPAC has been provided by national institutions, in particular the institutions participating in the {\it Gaia} Multilateral Agreement. This research has made use of the SIMBAD database, operated at CDS, Strasbourg, France. This research made use of Astropy (\url{http://www.astropy.org}), a community-developed core Python package for Astronomy.

\section*{Author contributions}

%Must include all authors, identified by initials, for example:

J.\ S.\ P.\ identified target, developed models of stellar environment, and calculated star-planet interaction flux density predictions. J.\ V.\ developed observing strategy, and reduced and analyzed radio data. Both authors contributed heavily to interpreting results and writing the manuscript.
%A.A. conceived the experiment(s),  A.A. and B.A. conducted the experiment(s), C.A. and D.A. analysed the results.  All authors reviewed the manuscript. 

\section*{Competing interests}

%To include, in this order: \textbf{Accession codes} (where applicable); \textbf{Competing interests} (mandatory statement). 

%The corresponding author is responsible for submitting a \href{http://www.nature.com/srep/policies/index.html#competing}{competing interests statement} on behalf of all authors of the paper. This statement must be included in the submitted article file.

The authors declare no competing interests. \\

\section*{Additional information}

\noindent Correspondence and additional information requests regarding this work can be directed to J.S.P.

The version of record of this article, first published in Nature Astronomy, is available online at Publisher’s website: \url{  https://www.nature.com/articles/s41550-023-01914-0 }.

%\clearpage 


\begin{thebibliography}{10}
\expandafter\ifx\csname url\endcsname\relax
  \def\url#1{\burl{#1}}\fi
\expandafter\ifx\csname urlprefix\endcsname\relax\def\urlprefix{URL }\fi
\providecommand{\bibinfo}[2]{#2}
\providecommand{\eprint}[2][]{\url{#2}}
\providecommand{\doi}[1]{\url{https://doi.org/#1}}
\bibcommenthead

\bibitem{Zarka2007}
\bibinfo{author}{{Zarka}, P.}
\newblock \bibinfo{title}{{Plasma interactions of exoplanets with their parent
  star and associated radio emissions}}.
\newblock \emph{\bibinfo{journal}{\planss}} \textbf{\bibinfo{volume}{55}},
  \bibinfo{pages}{598--617} (\bibinfo{year}{2007}).

\bibitem{Hallinan2013}
\bibinfo{author}{{Hallinan}, G.} \emph{et~al.}
\newblock \bibinfo{title}{{Looking for a Pulse: A Search for Rotationally
  Modulated Radio Emission from the Hot Jupiter, {\ensuremath{\tau}} Bo{\"o}tis
  b}}.
\newblock \emph{\bibinfo{journal}{\apj}} \textbf{\bibinfo{volume}{762}},
  \bibinfo{pages}{34} (\bibinfo{year}{2013}).

\bibitem{Pineda2018}
\bibinfo{author}{{Pineda}, J.~S.} \& \bibinfo{author}{{Hallinan}, G.}
\newblock \bibinfo{title}{{A Deep Radio Limit for the TRAPPIST-1 System}}.
\newblock \emph{\bibinfo{journal}{\apj}} \textbf{\bibinfo{volume}{866}},
  \bibinfo{pages}{155} (\bibinfo{year}{2018}).

\bibitem{Lazio2004}
\bibinfo{author}{{Lazio}, W., T.~Joseph} \emph{et~al.}
\newblock \bibinfo{title}{{The Radiometric Bode's Law and Extrasolar Planets}}.
\newblock \emph{\bibinfo{journal}{\apj}} \textbf{\bibinfo{volume}{612}},
  \bibinfo{pages}{511--518} (\bibinfo{year}{2004}).

\bibitem{Griessmeier2007}
\bibinfo{author}{{Grie{\ss}meier}, J.~M.}, \bibinfo{author}{{Zarka}, P.} \&
  \bibinfo{author}{{Spreeuw}, H.}
\newblock \bibinfo{title}{{Predicting low-frequency radio fluxes of known
  extrasolar planets}}.
\newblock \emph{\bibinfo{journal}{\aap}} \textbf{\bibinfo{volume}{475}},
  \bibinfo{pages}{359--368} (\bibinfo{year}{2007}).

\bibitem{Saur2013}
\bibinfo{author}{{Saur}, J.}, \bibinfo{author}{{Grambusch}, T.},
  \bibinfo{author}{{Duling}, S.}, \bibinfo{author}{{Neubauer}, F.~M.} \&
  \bibinfo{author}{{Simon}, S.}
\newblock \bibinfo{title}{{Magnetic energy fluxes in sub-Alfv{\'e}nic planet
  star and moon planet interactions}}.
\newblock \emph{\bibinfo{journal}{\aap}} \textbf{\bibinfo{volume}{552}},
  \bibinfo{pages}{A119} (\bibinfo{year}{2013}).

\bibitem{Turnpenney2018}
\bibinfo{author}{{Turnpenney}, S.}, \bibinfo{author}{{Nichols}, J.~D.},
  \bibinfo{author}{{Wynn}, G.~A.} \& \bibinfo{author}{{Burleigh}, M.~R.}
\newblock \bibinfo{title}{{Exoplanet-induced Radio Emission from M Dwarfs}}.
\newblock \emph{\bibinfo{journal}{\apj}} \textbf{\bibinfo{volume}{854}},
  \bibinfo{pages}{72} (\bibinfo{year}{2018}).

\bibitem{Queinnec1998JGR...10326649Q}
\bibinfo{author}{{Queinnec}, J.} \& \bibinfo{author}{{Zarka}, P.}
\newblock \bibinfo{title}{{Io-controlled decameter arcs and Io-Jupiter
  interaction}}.
\newblock \emph{\bibinfo{journal}{\jgr}} \textbf{\bibinfo{volume}{103}},
  \bibinfo{pages}{26649--26666} (\bibinfo{year}{1998}).

\bibitem{Vedantham2020}
\bibinfo{author}{{Vedantham}, H.~K.} \emph{et~al.}
\newblock \bibinfo{title}{{Coherent radio emission from a quiescent red dwarf
  indicative of star-planet interaction}}.
\newblock \emph{\bibinfo{journal}{Nature Astronomy}}  (\bibinfo{year}{2020}).

\bibitem{Callingham2021NatAs...5.1233C}
\bibinfo{author}{{Callingham}, J.~R.} \emph{et~al.}
\newblock \bibinfo{title}{{The population of M dwarfs observed at low radio
  frequencies}}.
\newblock \emph{\bibinfo{journal}{Nature Astronomy}}
  \textbf{\bibinfo{volume}{5}}, \bibinfo{pages}{1233--1239}
  (\bibinfo{year}{2021}).

\bibitem{Pope2021ApJ...919L..10P}
\bibinfo{author}{{Pope}, B. J.~S.} \emph{et~al.}
\newblock \bibinfo{title}{{The TESS View of LOFAR Radio-emitting Stars}}.
\newblock \emph{\bibinfo{journal}{\apjl}} \textbf{\bibinfo{volume}{919}},
  \bibinfo{pages}{L10} (\bibinfo{year}{2021}).

\bibitem{Pope2020}
\bibinfo{author}{{Pope}, B. J.~S.} \emph{et~al.}
\newblock \bibinfo{title}{{No Massive Companion to the Coherent Radio-emitting
  M Dwarf GJ 1151}}.
\newblock \emph{\bibinfo{journal}{\apjl}} \textbf{\bibinfo{volume}{890}},
  \bibinfo{pages}{L19} (\bibinfo{year}{2020}).

\bibitem{Perger2021}
\bibinfo{author}{{Perger}, M.} \emph{et~al.}
\newblock \bibinfo{title}{{The CARMENES search for exoplanets around M dwarfs.
  No evidence for a super-Earth in a 2-day orbit around GJ 1151}}.
\newblock \emph{\bibinfo{journal}{\aap}} \textbf{\bibinfo{volume}{649}},
  \bibinfo{pages}{L12} (\bibinfo{year}{2021}).

\bibitem{PerezTorres2021AA...645A..77P}
\bibinfo{author}{{P{\'e}rez-Torres}, M.} \emph{et~al.}
\newblock \bibinfo{title}{{Monitoring the radio emission of Proxima Centauri}}.
\newblock \emph{\bibinfo{journal}{\aap}} \textbf{\bibinfo{volume}{645}},
  \bibinfo{pages}{A77} (\bibinfo{year}{2021}).

\bibitem{Kavanagh2021MNRAS.504.1511K}
\bibinfo{author}{{Kavanagh}, R.~D.} \emph{et~al.}
\newblock \bibinfo{title}{{Planet-induced radio emission from the coronae of M
  dwarfs: the case of Prox Cen and AU Mic}}.
\newblock \emph{\bibinfo{journal}{\mnras}} \textbf{\bibinfo{volume}{504}},
  \bibinfo{pages}{1511--1518} (\bibinfo{year}{2021}).

\bibitem{Lynch2017}
\bibinfo{author}{{Lynch}, C.~R.}, \bibinfo{author}{{Lenc}, E.},
  \bibinfo{author}{{Kaplan}, D.~L.}, \bibinfo{author}{{Murphy}, T.} \&
  \bibinfo{author}{{Anderson}, G.~E.}
\newblock \bibinfo{title}{{154 MHz Detection of Faint, Polarized Flares from UV
  Ceti}}.
\newblock \emph{\bibinfo{journal}{\apjl}} \textbf{\bibinfo{volume}{836}},
  \bibinfo{pages}{L30} (\bibinfo{year}{2017}).

\bibitem{Villadsen2019}
\bibinfo{author}{{Villadsen}, J.} \& \bibinfo{author}{{Hallinan}, G.}
\newblock \bibinfo{title}{{Ultra-wideband Detection of 22 Coherent Radio Bursts
  on M Dwarfs}}.
\newblock \emph{\bibinfo{journal}{\apj}} \textbf{\bibinfo{volume}{871}},
  \bibinfo{pages}{214} (\bibinfo{year}{2019}).

\bibitem{Zic2020ApJ...905...23Z}
\bibinfo{author}{{Zic}, A.} \emph{et~al.}
\newblock \bibinfo{title}{{A Flare-type IV Burst Event from Proxima Centauri
  and Implications for Space Weather}}.
\newblock \emph{\bibinfo{journal}{\apj}} \textbf{\bibinfo{volume}{905}},
  \bibinfo{pages}{23} (\bibinfo{year}{2020}).

\bibitem{Perley2011}
\bibinfo{author}{{Perley}, R.~A.}, \bibinfo{author}{{Chandler}, C.~J.},
  \bibinfo{author}{{Butler}, B.~J.} \& \bibinfo{author}{{Wrobel}, J.~M.}
\newblock \bibinfo{title}{{The Expanded Very Large Array: A New Telescope for
  New Science}}.
\newblock \emph{\bibinfo{journal}{\apjl}} \textbf{\bibinfo{volume}{739}},
  \bibinfo{pages}{L1} (\bibinfo{year}{2011}).

\bibitem{AstudilloDefru2017}
\bibinfo{author}{{Astudillo-Defru}, N.} \emph{et~al.}
\newblock \bibinfo{title}{{The HARPS search for southern extra-solar planets.
  XLII. A system of Earth-mass planets around the nearby M dwarf YZ Ceti}}.
\newblock \emph{\bibinfo{journal}{\aap}} \textbf{\bibinfo{volume}{605}},
  \bibinfo{pages}{L11} (\bibinfo{year}{2017}).

\bibitem{Stock2020}
\bibinfo{author}{{Stock}, S.} \emph{et~al.}
\newblock \bibinfo{title}{{The CARMENES search for exoplanets around M dwarfs.
  Characterization of the nearby ultra-compact multiplanetary system YZ Ceti}}.
\newblock \emph{\bibinfo{journal}{\aap}} \textbf{\bibinfo{volume}{636}},
  \bibinfo{pages}{A119} (\bibinfo{year}{2020}).

\bibitem{Gudel2002ARAA..40..217G}
\bibinfo{author}{{G{\"u}del}, M.}
\newblock \bibinfo{title}{{Stellar Radio Astronomy: Probing Stellar Atmospheres
  from Protostars to Giants}}.
\newblock \emph{\bibinfo{journal}{\araa}} \textbf{\bibinfo{volume}{40}},
  \bibinfo{pages}{217--261} (\bibinfo{year}{2002}).

\bibitem{Dulk1985ARA&A..23..169D}
\bibinfo{author}{{Dulk}, G.~A.}
\newblock \bibinfo{title}{{Radio emission from the sun and stars.}}
\newblock \emph{\bibinfo{journal}{\araa}} \textbf{\bibinfo{volume}{23}},
  \bibinfo{pages}{169--224} (\bibinfo{year}{1985}).

\bibitem{Osten2008ApJ...674.1078O}
\bibinfo{author}{{Osten}, R.~A.} \& \bibinfo{author}{{Bastian}, T.~S.}
\newblock \bibinfo{title}{{Ultrahigh Time Resolution Observations of Radio
  Bursts on AD Leonis}}.
\newblock \emph{\bibinfo{journal}{\apj}} \textbf{\bibinfo{volume}{674}},
  \bibinfo{pages}{1078--1085} (\bibinfo{year}{2008}).

\bibitem{Lanza2009}
\bibinfo{author}{{Lanza}, A.~F.}
\newblock \bibinfo{title}{{Stellar coronal magnetic fields and star-planet
  interaction}}.
\newblock \emph{\bibinfo{journal}{\aap}} \textbf{\bibinfo{volume}{505}},
  \bibinfo{pages}{339--350} (\bibinfo{year}{2009}).

\bibitem{Zarka2018A&A...618A..84Z}
\bibinfo{author}{{Zarka}, P.} \emph{et~al.}
\newblock \bibinfo{title}{{Jupiter radio emission induced by Ganymede and
  consequences for the radio detection of exoplanets}}.
\newblock \emph{\bibinfo{journal}{\aap}} \textbf{\bibinfo{volume}{618}},
  \bibinfo{pages}{A84} (\bibinfo{year}{2018}).

\bibitem{Fischer2019ApJ...872..113F}
\bibinfo{author}{{Fischer}, C.} \& \bibinfo{author}{{Saur}, J.}
\newblock \bibinfo{title}{{Time-variable Electromagnetic Star-Planet
  Interaction: The TRAPPIST-1 System as an Exemplary Case}}.
\newblock \emph{\bibinfo{journal}{\apj}} \textbf{\bibinfo{volume}{872}},
  \bibinfo{pages}{113} (\bibinfo{year}{2019}).

\bibitem{Hess2011AA...531A..29H}
\bibinfo{author}{{Hess}, S.~L.~G.} \& \bibinfo{author}{{Zarka}, P.}
\newblock \bibinfo{title}{{Modeling the radio signature of the orbital
  parameters, rotation, and magnetic field of exoplanets}}.
\newblock \emph{\bibinfo{journal}{\aap}} \textbf{\bibinfo{volume}{531}},
  \bibinfo{pages}{A29} (\bibinfo{year}{2011}).

\bibitem{Newton2017ApJ...834...85N}
\bibinfo{author}{{Newton}, E.~R.} \emph{et~al.}
\newblock \bibinfo{title}{{The H{\ensuremath{\alpha}} Emission of Nearby M
  Dwarfs and its Relation to Stellar Rotation}}.
\newblock \emph{\bibinfo{journal}{\apj}} \textbf{\bibinfo{volume}{834}},
  \bibinfo{pages}{85} (\bibinfo{year}{2017}).

\bibitem{Nichols2012ApJ...760...59N}
\bibinfo{author}{{Nichols}, J.~D.} \emph{et~al.}
\newblock \bibinfo{title}{{Origin of Electron Cyclotron Maser Induced Radio
  Emissions at Ultracool Dwarfs: Magnetosphere-Ionosphere Coupling Currents}}.
\newblock \emph{\bibinfo{journal}{\apj}} \textbf{\bibinfo{volume}{760}},
  \bibinfo{pages}{59} (\bibinfo{year}{2012}).

\bibitem{Loyd2018ApJ...867...71L}
\bibinfo{author}{{Loyd}, R.~O.~P.} \emph{et~al.}
\newblock \bibinfo{title}{{The MUSCLES Treasury Survey. V. FUV Flares on Active
  and Inactive M Dwarfs}}.
\newblock \emph{\bibinfo{journal}{\apj}} \textbf{\bibinfo{volume}{867}},
  \bibinfo{pages}{71} (\bibinfo{year}{2018}).

\bibitem{Slee2003}
\bibinfo{author}{{Slee}, O.~B.}, \bibinfo{author}{{Willes}, A.~J.} \&
  \bibinfo{author}{{Robinson}, R.~D.}
\newblock \bibinfo{title}{{Long-duration Coherent Radio Emission from the dMe
  Star Proxima Centauri}}.
\newblock \emph{\bibinfo{journal}{\pasa}} \textbf{\bibinfo{volume}{20}},
  \bibinfo{pages}{257--262} (\bibinfo{year}{2003}).

\bibitem{Pineda2021ApJ...918...40P}
\bibinfo{author}{{Pineda}, J.~S.}, \bibinfo{author}{{Youngblood}, A.} \&
  \bibinfo{author}{{France}, K.}
\newblock \bibinfo{title}{{The M-dwarf Ultraviolet Spectroscopic Sample. I.
  Determining Stellar Parameters for Field Stars}}.
\newblock \emph{\bibinfo{journal}{\apj}} \textbf{\bibinfo{volume}{918}},
  \bibinfo{pages}{40} (\bibinfo{year}{2021}).

\bibitem{Gaia2018A&A...616A...1G}
\bibinfo{author}{{Gaia Collaboration}} \emph{et~al.}
\newblock \bibinfo{title}{{Gaia Data Release 2. Summary of the contents and
  survey properties}}.
\newblock \emph{\bibinfo{journal}{\aap}} \textbf{\bibinfo{volume}{616}},
  \bibinfo{pages}{A1} (\bibinfo{year}{2018}).

\bibitem{Mann2015}
\bibinfo{author}{{Mann}, A.~W.}, \bibinfo{author}{{Feiden}, G.~A.},
  \bibinfo{author}{{Gaidos}, E.}, \bibinfo{author}{{Boyajian}, T.} \&
  \bibinfo{author}{{von Braun}, K.}
\newblock \bibinfo{title}{{How to Constrain Your M Dwarf: Measuring Effective
  Temperature, Bolometric Luminosity, Mass, and Radius}}.
\newblock \emph{\bibinfo{journal}{\apj}} \textbf{\bibinfo{volume}{804}},
  \bibinfo{pages}{64} (\bibinfo{year}{2015}).

\bibitem{Boyajian2012}
\bibinfo{author}{{Boyajian}, T.~S.} \emph{et~al.}
\newblock \bibinfo{title}{{Stellar Diameters and Temperatures. II.
  Main-sequence K- and M-stars}}.
\newblock \emph{\bibinfo{journal}{\apj}} \textbf{\bibinfo{volume}{757}},
  \bibinfo{pages}{112} (\bibinfo{year}{2012}).

\bibitem{Ricker2015}
\bibinfo{author}{{Ricker}, G.~R.} \emph{et~al.}
\newblock \bibinfo{title}{{Transiting Exoplanet Survey Satellite (TESS)}}.
\newblock \emph{\bibinfo{journal}{Journal of Astronomical Telescopes,
  Instruments, and Systems}} \textbf{\bibinfo{volume}{1}},
  \bibinfo{pages}{014003} (\bibinfo{year}{2015}).

\bibitem{mcmullin2007}
\bibinfo{author}{{McMullin}, J.~P.}, \bibinfo{author}{{Waters}, B.},
  \bibinfo{author}{{Schiebel}, D.}, \bibinfo{author}{{Young}, W.} \&
  \bibinfo{author}{{Golap}, K.}
\newblock \bibinfo{editor}{{Shaw}, R.~A.}, \bibinfo{editor}{{Hill}, F.} \&
  \bibinfo{editor}{{Bell}, D.~J.} (eds) \emph{\bibinfo{title}{{CASA
  Architecture and Applications}}}.
\newblock (eds \bibinfo{editor}{{Shaw}, R.~A.}, \bibinfo{editor}{{Hill}, F.} \&
  \bibinfo{editor}{{Bell}, D.~J.}) \emph{\bibinfo{booktitle}{Astronomical Data
  Analysis Software and Systems XVI}}, Vol. \bibinfo{volume}{376} of
  \emph{\bibinfo{series}{Astronomical Society of the Pacific Conference
  Series}}, \bibinfo{pages}{127} (\bibinfo{year}{2007}).

\bibitem{Benz1994A&A...285..621B}
\bibinfo{author}{{Benz}, A.~O.} \& \bibinfo{author}{{Guedel}, M.}
\newblock \bibinfo{title}{{X-ray/microwave ratio of flares and coronae}}.
\newblock \emph{\bibinfo{journal}{\aap}} \textbf{\bibinfo{volume}{285}},
  \bibinfo{pages}{621--630} (\bibinfo{year}{1994}).

\bibitem{See2017}
\bibinfo{author}{{See}, V.} \emph{et~al.}
\newblock \bibinfo{title}{{Studying stellar spin-down with Zeeman-Doppler
  magnetograms}}.
\newblock \emph{\bibinfo{journal}{\mnras}} \textbf{\bibinfo{volume}{466}},
  \bibinfo{pages}{1542--1554} (\bibinfo{year}{2017}).

\bibitem{Weber1967}
\bibinfo{author}{{Weber}, E.~J.} \& \bibinfo{author}{{Davis}, J., Leverett}.
\newblock \bibinfo{title}{{The Angular Momentum of the Solar Wind}}.
\newblock \emph{\bibinfo{journal}{\apj}} \textbf{\bibinfo{volume}{148}},
  \bibinfo{pages}{217--227} (\bibinfo{year}{1967}).

\bibitem{Parker1958}
\bibinfo{author}{{Parker}, E.~N.}
\newblock \bibinfo{title}{{Dynamics of the Interplanetary Gas and Magnetic
  Fields.}}
\newblock \emph{\bibinfo{journal}{\apj}} \textbf{\bibinfo{volume}{128}},
  \bibinfo{pages}{664} (\bibinfo{year}{1958}).

\bibitem{Stelzer2013}
\bibinfo{author}{{Stelzer}, B.}, \bibinfo{author}{{Marino}, A.},
  \bibinfo{author}{{Micela}, G.}, \bibinfo{author}{{L{\'o}pez-Santiago}, J.} \&
  \bibinfo{author}{{Liefke}, C.}
\newblock \bibinfo{title}{{The UV and X-ray activity of the M dwarfs within 10
  pc of the Sun}}.
\newblock \emph{\bibinfo{journal}{\mnras}} \textbf{\bibinfo{volume}{431}},
  \bibinfo{pages}{2063--2079} (\bibinfo{year}{2013}).

\bibitem{Loyd2016ApJ...824..102L}
\bibinfo{author}{{Loyd}, R.~O.~P.} \emph{et~al.}
\newblock \bibinfo{title}{{The MUSCLES Treasury Survey. III. X-Ray to Infrared
  Spectra of 11 M and K Stars Hosting Planets}}.
\newblock \emph{\bibinfo{journal}{\apj}} \textbf{\bibinfo{volume}{824}},
  \bibinfo{pages}{102} (\bibinfo{year}{2016}).

\bibitem{Klein2021MNRAS.500.1844K}
\bibinfo{author}{{Klein}, B.} \emph{et~al.}
\newblock \bibinfo{title}{{The large-scale magnetic field of Proxima Centauri
  near activity maximum}}.
\newblock \emph{\bibinfo{journal}{\mnras}} \textbf{\bibinfo{volume}{500}},
  \bibinfo{pages}{1844--1850} (\bibinfo{year}{2021}).

\bibitem{Moutou2017}
\bibinfo{author}{{Moutou}, C.} \emph{et~al.}
\newblock \bibinfo{title}{{SPIRou input catalogue: activity, rotation and
  magnetic field of cool dwarfs}}.
\newblock \emph{\bibinfo{journal}{\mnras}} \textbf{\bibinfo{volume}{472}},
  \bibinfo{pages}{4563--4586} (\bibinfo{year}{2017}).

\bibitem{Morin2010MNRAS.407.2269M}
\bibinfo{author}{{Morin}, J.} \emph{et~al.}
\newblock \bibinfo{title}{{Large-scale magnetic topologies of late M dwarfs*}}.
\newblock \emph{\bibinfo{journal}{\mnras}} \textbf{\bibinfo{volume}{407}},
  \bibinfo{pages}{2269--2286} (\bibinfo{year}{2010}).

\bibitem{Reiners2022A&A...662A..41R}
\bibinfo{author}{{Reiners}, A.} \emph{et~al.}
\newblock \bibinfo{title}{{Magnetism, rotation, and nonthermal emission in cool
  stars. Average magnetic field measurements in 292 M dwarfs}}.
\newblock \emph{\bibinfo{journal}{\aap}} \textbf{\bibinfo{volume}{662}},
  \bibinfo{pages}{A41} (\bibinfo{year}{2022}).

\bibitem{Vidotto2014MNRAS.438.1162V}
\bibinfo{author}{{Vidotto}, A.~A.} \emph{et~al.}
\newblock \bibinfo{title}{{M-dwarf stellar winds: the effects of realistic
  magnetic geometry on rotational evolution and planets}}.
\newblock \emph{\bibinfo{journal}{\mnras}} \textbf{\bibinfo{volume}{438}},
  \bibinfo{pages}{1162--1175} (\bibinfo{year}{2014}).

\bibitem{Altschuler1969SoPh....9..131A}
\bibinfo{author}{{Altschuler}, M.~D.} \& \bibinfo{author}{{Newkirk}, G.}
\newblock \bibinfo{title}{{Magnetic Fields and the Structure of the Solar
  Corona. I: Methods of Calculating Coronal Fields}}.
\newblock \emph{\bibinfo{journal}{\solphys}} \textbf{\bibinfo{volume}{9}},
  \bibinfo{pages}{131--149} (\bibinfo{year}{1969}).

\bibitem{Wood2001ApJ...547L..49W}
\bibinfo{author}{{Wood}, B.~E.}, \bibinfo{author}{{Linsky}, J.~L.},
  \bibinfo{author}{{M{\"u}ller}, H.-R.} \& \bibinfo{author}{{Zank}, G.~P.}
\newblock \bibinfo{title}{{Observational Estimates for the Mass-Loss Rates of
  {\ensuremath{\alpha}} Centauri and Proxima Centauri Using Hubble Space
  Telescope Ly{\ensuremath{\alpha}} Spectra}}.
\newblock \emph{\bibinfo{journal}{\apjl}} \textbf{\bibinfo{volume}{547}},
  \bibinfo{pages}{L49--L52} (\bibinfo{year}{2001}).

\bibitem{Wood2021ApJ...915...37W}
\bibinfo{author}{{Wood}, B.~E.} \emph{et~al.}
\newblock \bibinfo{title}{{New Observational Constraints on the Winds of M
  dwarf Stars}}.
\newblock \emph{\bibinfo{journal}{\apj}} \textbf{\bibinfo{volume}{915}},
  \bibinfo{pages}{37} (\bibinfo{year}{2021}).

\bibitem{Vidotto2021LRSP...18....3V}
\bibinfo{author}{{Vidotto}, A.~A.}
\newblock \bibinfo{title}{{The evolution of the solar wind}}.
\newblock \emph{\bibinfo{journal}{Living Reviews in Solar Physics}}
  \textbf{\bibinfo{volume}{18}}, \bibinfo{pages}{3} (\bibinfo{year}{2021}).

\bibitem{Queinnec2001P&SS...49..365Q}
\bibinfo{author}{{Queinnec}, J.} \& \bibinfo{author}{{Zarka}, P.}
\newblock \bibinfo{title}{{Flux, power, energy and polarization of Jovian
  S-bursts}}.
\newblock \emph{\bibinfo{journal}{\planss}} \textbf{\bibinfo{volume}{49}},
  \bibinfo{pages}{365--376} (\bibinfo{year}{2001}).

\bibitem{astropy2013A&A...558A..33A}
\bibinfo{author}{{Astropy Collaboration}} \emph{et~al.}
\newblock \bibinfo{title}{{Astropy: A community Python package for astronomy}}.
\newblock \emph{\bibinfo{journal}{\aap}} \textbf{\bibinfo{volume}{558}},
  \bibinfo{pages}{A33} (\bibinfo{year}{2013}).

\bibitem{astropy2018AJ....156..123A}
\bibinfo{author}{{Astropy Collaboration}} \emph{et~al.}
\newblock \bibinfo{title}{{The Astropy Project: Building an Open-science
  Project and Status of the v2.0 Core Package}}.
\newblock \emph{\bibinfo{journal}{\aj}} \textbf{\bibinfo{volume}{156}},
  \bibinfo{pages}{123} (\bibinfo{year}{2018}).

\bibitem{Reid1995}
\bibinfo{author}{{Reid}, I.~N.}, \bibinfo{author}{{Hawley}, S.~L.} \&
  \bibinfo{author}{{Gizis}, J.~E.}
\newblock \bibinfo{title}{{The Palomar/MSU Nearby-Star Spectroscopic Survey. I.
  The Northern M Dwarfs -Bandstrengths and Kinematics}}.
\newblock \emph{\bibinfo{journal}{\aj}} \textbf{\bibinfo{volume}{110}},
  \bibinfo{pages}{1838} (\bibinfo{year}{1995}).

\bibitem{Hawley1996}
\bibinfo{author}{{Hawley}, S.~L.}, \bibinfo{author}{{Gizis}, J.~E.} \&
  \bibinfo{author}{{Reid}, I.~N.}
\newblock \bibinfo{title}{{The Palomar/MSU Nearby Star Spectroscopic
  Survey.II.The Southern M Dwarfs and Investigation of Magnetic Activity}}.
\newblock \emph{\bibinfo{journal}{\aj}} \textbf{\bibinfo{volume}{112}},
  \bibinfo{pages}{2799} (\bibinfo{year}{1996}).

\bibitem{Wargelin2017MNRAS.464.3281W}
\bibinfo{author}{{Wargelin}, B.~J.}, \bibinfo{author}{{Saar}, S.~H.},
  \bibinfo{author}{{Pojma{\'n}ski}, G.}, \bibinfo{author}{{Drake}, J.~J.} \&
  \bibinfo{author}{{Kashyap}, V.~L.}
\newblock \bibinfo{title}{{Optical, UV, and X-ray evidence for a 7-yr stellar
  cycle in Proxima Centauri}}.
\newblock \emph{\bibinfo{journal}{\mnras}} \textbf{\bibinfo{volume}{464}},
  \bibinfo{pages}{3281--3296} (\bibinfo{year}{2017}).

\bibitem{Reiners2018}
\bibinfo{author}{{Reiners}, A.} \emph{et~al.}
\newblock \bibinfo{title}{{The CARMENES search for exoplanets around M dwarfs.
  High-resolution optical and near-infrared spectroscopy of 324 survey stars}}.
\newblock \emph{\bibinfo{journal}{\aap}} \textbf{\bibinfo{volume}{612}},
  \bibinfo{pages}{A49} (\bibinfo{year}{2018}).

\bibitem{Newton2018}
\bibinfo{author}{{Newton}, E.~R.}, \bibinfo{author}{{Mondrik}, N.},
  \bibinfo{author}{{Irwin}, J.}, \bibinfo{author}{{Winters}, J.~G.} \&
  \bibinfo{author}{{Charbonneau}, D.}
\newblock \bibinfo{title}{{New Rotation Period Measurements for M Dwarfs in the
  Southern Hemisphere: An Abundance of Slowly Rotating, Fully Convective
  Stars}}.
\newblock \emph{\bibinfo{journal}{\aj}} \textbf{\bibinfo{volume}{156}},
  \bibinfo{pages}{217} (\bibinfo{year}{2018}).

\bibitem{Irwin2011}
\bibinfo{author}{{Irwin}, J.} \emph{et~al.}
\newblock \bibinfo{title}{{On the Angular Momentum Evolution of Fully
  Convective Stars: Rotation Periods for Field M-dwarfs from the MEarth Transit
  Survey}}.
\newblock \emph{\bibinfo{journal}{\apj}} \textbf{\bibinfo{volume}{727}},
  \bibinfo{pages}{56} (\bibinfo{year}{2011}).

\bibitem{Reiners2008b}
\bibinfo{author}{{Reiners}, A.} \& \bibinfo{author}{{Basri}, G.}
\newblock \bibinfo{title}{{The moderate magnetic field of the flare star
  Proxima Centauri}}.
\newblock \emph{\bibinfo{journal}{\aap}} \textbf{\bibinfo{volume}{489}},
  \bibinfo{pages}{L45--L48} (\bibinfo{year}{2008}).

\bibitem{Faria2022AA...658A.115F}
\bibinfo{author}{{Faria}, J.~P.} \emph{et~al.}
\newblock \bibinfo{title}{{A candidate short-period sub-Earth orbiting Proxima
  Centauri}}.
\newblock \emph{\bibinfo{journal}{\aap}} \textbf{\bibinfo{volume}{658}},
  \bibinfo{pages}{A115} (\bibinfo{year}{2022}).

\bibitem{AngladaEscude2016}
\bibinfo{author}{{Anglada-Escud{\'e}}, G.} \emph{et~al.}
\newblock \bibinfo{title}{{A terrestrial planet candidate in a temperate orbit
  around Proxima Centauri}}.
\newblock \emph{\bibinfo{journal}{\nat}} \textbf{\bibinfo{volume}{536}},
  \bibinfo{pages}{437--440} (\bibinfo{year}{2016}).

\end{thebibliography}
\end{document}